\documentclass[fleqn,usenatbib]{mnras}

\usepackage{newtxtext,newtxmath}
\usepackage{threeparttable}

\usepackage[T1]{fontenc}

\DeclareRobustCommand{\VAN}[3]{#2}
\let\VANthebibliography\thebibliography
\def\thebibliography{\DeclareRobustCommand{\VAN}[3]{##3}\VANthebibliography}

\usepackage{graphicx}	
\usepackage{amsmath}	

\title[Not so peculiar?]{An unexplored enrichment stochasticity and its implications for stellar abundance patterns}

\author[A. Aggarwal et al.]{
Anmol Aggarwal,$^{1}$\thanks{E-mail: anmol.aggarwal.24@ucl.ac.uk}
Ralph Sch\"{o}nrich,$^{1}$
\\
$^{1}$Mullard Space Science Laboratory, University College London, Holmbury St. Mary, Dorking, RH5 6NT, Surrey, UK\\
}

\date{Accepted XXX. Received YYY; in original form ZZZ}

\pubyear{\the\year{}}

\begin{document}
\label{firstpage}
\pagerange{\pageref{firstpage}--\pageref{lastpage}}
\maketitle

\begin{abstract}
Extremely low metallicity stars are intensely studied as they take observations the closest to the very first generations of stars in the universe. Widely assumed to be enriched by just one dying massive star, some of these very metal poor stars have abnormal chemical abundance ratios and have been taken to reflect a rare hypernova (with high explosion energy $\gtrsim$ 10\textsuperscript{52} erg.). Here we remodel the enrichment of three such stars and show that their abundances are better explained by enrichment from a normal (less energetic) supernova accounting for inhomogeneous distribution of the ejecta. This work establishes the importance of the inhomogeneity of supernovae, serves as a template for a required reassessment of all metal-poor/peculiar stars, and raises the need to quantify this inhomogeneity both in theory and in observations.
\end{abstract}




\section{Introduction}

The chemical elements present in low-mass stars largely reflect the composition with which those stars have been born (apart from, e.g., elements carried away in stellar winds \citep{deutsch1956circumstellar, cassinelli1979stellar}, or affected by stellar gravitational settling \citep{proffitt1991gravitational, stancliffe2008thermohaline, Korn_2007}. This way, stars convey information about the circumstances and place of their birth.

A particular focus has been on charting the very first stars to understand the conditions of early galaxies. It has been hypothesised that this first generation of stars was metal-free, yet the hunt for such stars has been uneventful till now, as they are rare, and many authors expect that the initial mass function is biased towards heavy, short-lived stars (\citep{tinsley1979stellar, Ohkubo_2009}, for modern opposing viewpoints, see e.g. \cite{Hutter2025} vs. \cite{Hartwig2015} and \cite{tacchella}). We note that for most stellar populations, low metallicity does not imply that they came before the more metal-rich objects \citep{white1999}, but rather encodes their origin in galactic outskirts or low-mass galaxies (dwarf galaxies until today have metallicities around [Fe/H] $\sim -2$; it is a fair assumption that below this level, metallicity does start to encode a very early origin of the star).

In the realm of understanding peculiar abundance patterns, there are plenty of approaches, in particular, stochastic chemical evolution models \citep{karlsson2005, 10.1093/mnras/stab281}. These commonly keep a locally well-mixed interstellar medium (ISM), but relinquish the common assumption of large number statistics in supernova (SN) yields in favour of single events and localised enrichment. We identify three distinct principal sources of stochasticity in chemical evolution:
\begin{enumerate}
    \item {\bf Stochasticity in galactic evolution/"evolution stochasticity"}. It is caused by a cosmologically motivated infall, where discrete accretion events in time drive spikes in star formation rate and thus loops in, e.g., the [Fe/H]-[$\alpha$/Fe] metallicity plane \citep{colavitti2008}.
    \item {\bf Stochasticity by low numbers/"Poisson stochasticity"}. Considering relatively small-sized systems that act as well-mixed bubbles, in which the Poisson noise of stellar yields and particularly SN explosions leads to fluctuations in the ISM composition \citep{karlsson2005, Koch2008, Venn, 10.1093/mnras/stab281}. Essentially, Poisson noise becomes important when the yields in a single element of a single event become comparable to the system's total content times the observational error (i.e. $\rm{[X/H]} \times \sigma_{\rm X}$).
    \item {\bf Stochasticity by incomplete mixing/"mixing stochasticity"}. At the time of its explosion, a SN is not well-mixed and, due to explosion non-uniformity, will expel different layers in different directions. This material will hit nearby star-forming clouds and so transfer fractions of the SN rather than a fraction of the entire mixed SN to the fresh star formation site. This third source of stochasticity has, to our knowledge, not been modelled. Here, parts of a SN can enrich different regions of the system's star-forming ISM/clouds.
\end{enumerate}

A growing number of studies circumnavigate the complications arising from points 1) and 2) by selecting very metal-poor, peculiar stars, which are then argued to have been enriched by a single SN event. However, this approach does tacitly assume that the elemental yields have to arrive as a well-mixed homogenous mixture at the star-forming cloud, i.e., either mixed perfectly within the explosion or in its dispersal. There is ample evidence (see below) that a supernova explosion distributes the elements non-uniformly to its surroundings. In their picture, the peculiar abundances of an observed star then imply a peculiar SN event – in some cases a hypernova (with energy $E_{\rm SN} \gtrsim 10^{52} {\rm erg}$). Here, we examine three such stars (AS0039 \citep{Skúladóttir_2021, Skúladóttir_2024}, HE 1327-2326 \citep{ezzeddine2019evidence}, and J0931+0038 \citep{ji2024spectacular}) where this conclusion has been drawn. These are three of the four stars claimed to be descendants of hypernovae in the present literature. The fourth, SMSS J200322.54-114203.3 \citep{yong2021r}, needs a separate discussion of its own as it involves r-process enrichment as well. We will tackle this claim in a subsequent paper.

We show an alternative argument here: instead of demanding a peculiar supernova, we consider point 3) above and allow for non-uniform distribution of the yields of a "normal" (standard model, $E_{\rm SN} \sim 10^{51} {\rm erg}$) supernova. This uneven distribution seems more natural, given that such inhomogeneity has been observed in recent supernova remnants and predicted from models \citep{Hughes_2000, Fesen_2006, Larsson_2013, meyer2015asymmetric, vartanyan2019successful, muller2020hydrodynamics, van2023modelling, vartanyan20253d}.

Here, our idea is to conclusively show the importance of the non-uniform dispersal of elements by a supernova progenitor rather than finding the best fit for each star. The fits only represent a possible outcome of taking such a process into account and necessitate to rethink previous conclusions for such stars. In the following, we describe the three stars and data used in Sections 2-4. In Section 5 we discuss other evidence and outline issues with the current approaches. In Section 6 we outline the fitting and apply it to the three stars. Here we make sure to fully use all abundances as used in the original papers to exclude the possibility of our conclusions being affected by different assumptions/abundances. In section 7 we discuss the results, and in Section 8 we conclude.

\section{The AS0039 star}
\label{as0039}

AS0039 \citep{Skúladóttir_2021, Skúladóttir_2024} is one of the most prominent peculiar and low-metallicity stars, a member of the Sculptor dwarf spheroidal galaxy \citep{1938BHarO.908....1S}. As detailed above, AS0039 was presented as a star enriched by one metal-free hypernova \citep{Skúladóttir_2021, iwamoto1998hypernova}. They fit to the observed abundances a model \citep{Heger_2010} with progenitor mass $21 {\rm M}_\odot$ and $E_{\rm SN} = 10^{52} {\rm \ erg}$ using the "Starfit" tool \footnote{https://starfit.org} (the Starfit tool can match the observed abundances of a star with a model from a database or with two or more models), and a $\sim (20 \pm 2) {\rm M}_\odot$ progenitor model using their own fitting routine \citep{10.1093/mnras/stz1464}. They reiterate these conclusions on updated chemical abundances \citep{Skúladóttir_2024}. For convenience, we report the non-(local thermal equilibrium) (NLTE) abundances from their Table 4 \citep{Skúladóttir_2024} in our Table \ref{tab:tab_1}. To conform with them, we use these values in combination with the same set of solar abundances \citep{asplund2021chemical}.

\begin{table}
    \centering
    \caption{NLTE chemical abundances of AS0039 \citep{Skúladóttir_2024}.}
    
    \begin{threeparttable}
    \begin{tabular}{ccc}
    
      \hline
      \hline
      Element & \text{[X/Fe]\textsubscript{NLTE}} & $\sigma$\\
      \hline
      Fe$^\dagger$  & -3.64 & 0.11\\
      C    & -0.59 & 0.20\\
      Na   & -0.53 & 0.07\\
      Mg   & +0.15 & 0.05\\
      Al   & +0.05 & 0.10\\
      Si   & -0.35 & 0.15\\
      Ca   & +0.08 & 0.11\\
      Sc   & -0.36 & 0.08\\
      Ti   & +0.40 & 0.06\\
      Cr   & +0.29 & 0.13\\
      Mn   & -0.04 & 0.23\\
      Co   & +0.59 & 0.08\\
      Ni   & +0.19 & 0.09\\
      Zn   & +0.63 & 0.30\\
      \hline

    \end{tabular}
    \begin{tablenotes}
    \footnotesize
    \item $^\dagger$[Fe/H]
    \item In our notation: "$\sigma$ (uncertainty)" is $\delta_{[X/Fe]}$ in \cite{Skúladóttir_2024}.
    \end{tablenotes}
    \end{threeparttable}
    \label{tab:tab_1}
\end{table}

Some abundance ratios are particularly stressed as the footprints of a hypernova: in particular, they emphasise the low [C/Fe] value, as lower-energy Pop III supernovae eject relatively less iron.

They also compare AS0039 and other metal-poor Sculptor stars to pre-existing measurements of other metal-poor Milky Way (MW) and ultra-faint dwarf (UFD) galaxies' stars, stating that metal-poor Sculptor stars have generally low [X/Fe] ratios for light elements (Z $\lesssim 20$) and are similar to MW and UFD stars for heavier elements. They posit that for heavier elements [X/Fe] ratios are less affected by explosion energy, while [X/Fe] ratios for light elements are anti-correlated to increasing explosion energy, arguing consistency with the hypernova picture.

\section{HE 1327-2326}

The Carbon Enhanced Metal Poor (in this case [Fe/H] = -5.20) star HE 1327-2326 has been attributed to hypernova enrichment \citep{ezzeddine2019evidence}. Their estimate for the explosion energy ($5 \times 10^{51}$ erg) is high but slightly below more traditional hypernova definitions ($E_{\rm SN} \gtrsim 10^{52}$ erg), i.e., an order of magnitude more than a classical supernova. Their leading argument for high explosion energy is enhanced zinc abundance ([Zn/Fe] $= 0.80 \pm 0.25$). The authors also require an aspherical hypernova explosion with bipolar jets to account for several data points. \cite{ezzeddine2019evidence} use various models with different parameters from \cite{Tominaga_2007}. Among those, the spherical models do not produce high enough [Sc/Fe], [Ti/Fe], and [Zn/Fe]; in particular, the high observed [Ti/Fe] convinces them to rule out spherical explosion models. They also argue that fast-rotating stars will naturally result in such an aspherical explosion and claim that HE 1327-2326's high [N/Fe] independently implies a fast-rotating progenitor. We remark that \cite{ezzeddine2019evidence} also noted the problem that the minihalo they posit as host of these events would not withstand a hypernova explosion \citep{cooke2014carbon}, where they then suggest that the enrichment might stem from a neighbouring system.

\begin{table}
    \centering
    \caption{Chemical abundances of HE 1327-2326 \citep{ezzeddine2019evidence}.}
    \begin{threeparttable}
    \begin{tabular}{l c c c c c}
        \hline
        \hline
        Element & $N_\text{lines}$ & $\mathrm{log}\epsilon (X)$ & $\sigma$ & [X/H] & [X/Fe]\\
        \hline
        C (CH) & syn. & 6.21 & 0.10 & -2.22 & 3.49\\
        N (NH) & syn. & 6.10 & 0.20 & -1.73 & 3.98\\
        O (OH) & syn. & 6.12 & 0.20 & -2.57 & 3.14\\
        Na I & 2 & 2.99 & 0.04 & -3.25 & 2.46\\
        Mg I & 4 & 3.54 & 0.02 & -4.06 & 1.65\\
        Al I & 1 & 1.90 & 0.03 & -4.55 & 1.16\\
        Si I & 1 & 2.80 & 0.27 & -4.71 & 1.28\\
        Ca II & 4 & 1.34 & 0.15 & -5.00 & 0.71\\
        Sc II & $\cdots$ & $<$-1.68 & $\cdots$ & $<$-4.83 & $<$0.88\\
        Ti II & 15 & -0.09 & 0.17 & -5.04 & 0.67\\
        Mn I & $\cdots$ & $<$0.53 & $\cdots$ & $<$-4.90 & $<$0.81\\
        Fe I & 10 & 1.79 & 0.15 & -5.71 & $\cdots$\\
        Fe I & 10 & 2.30 & 0.11 & -5.20 & $\cdots$\\
        Fe II & 4 & 1.51 & 0.26 & -5.99 & $\cdots$\\
        Co I & $\cdots$ & $<$0.58 & $\cdots$ & $<$-4.41 & $<$1.30\\
        Ni I & 4 & 0.73 & 0.20 & -5.49 & 0.33\\
        Zn I & 1 & 0.16 & 0.25 & -4.40 & 0.80\\
        Sr II & 2 & -1.76 & 0.06 & -4.63 & 1.08\\
        \hline
    \end{tabular}

    \begin{tablenotes}
    \footnotesize
    \item "In our notation: $\sigma$ (uncertainty)" is $\sigma \mathrm{log}\epsilon (X)$ in \cite{ezzeddine2019evidence}
    \end{tablenotes}
    
    \end{threeparttable}
    \label{tab:1}
\end{table}

In Table \ref{tab:1} we list the elemental abundance values from \cite{ezzeddine2019evidence}, both their [X/H] and [X/Fe], as they use different [Fe/H] to translate different elements. For Sc, Mn, and Co there are only upper limits. For consistency with \cite{ezzeddine2019evidence}, we use for this star solar values from \cite{asplund2009chemical}.

For Ni, \cite{ezzeddine2019evidence} use an [Fe/H] of -5.82, which is not listed in the table.

\section{J0931+0038}

2MASSJ09311004+0038042 (J0931+0038) has a relatively higher metallicity at [Fe/H] = -1.76, but the claim about the progenitor of J0931+0038 is more extreme than AS0039 and HE 1327-2326. This star's abundance pattern has been proposed to originate from a $>$ 50 \(M_{\odot}\) star, asserting as the best option an 80 \(M_{\odot}\) progenitor with, even for a hypernova, an extraordinary explosion energy of $2.2 \times 10^{52}$ erg \citep{ji2024spectacular}.

To support such an extraordinary (in the authors' own words) claim, \cite{ji2024spectacular} first point out that despite the comparatively high metallicity, the star has low abundances of Na, K, Sc, and Ba, which is a sign of enrichment from a single, dominant source. Second, the abundances of light elements (C to Sc) show a strong odd-even effect, which they use to draw parallels with another star claimed to be a descendant of a massive pair-instability supernova (PISN) \citep{xing2023metal}. We note that the star reported by \cite{xing2023metal} has been reanalysed by other groups, and their claim of PISN enrichment has been refuted \citep{Thibodeaux2024LAMOST, jeena2024core, Jeena2024Origin, skuladottir2024pair}. In the spirit of our above discussions, we remark that also the reanalyses do not take into account the asymmetric dispersal of yields, which is our topic here, so the whole discussion needs to be revisited with our toolset. Third, the abundances of Sc, Ti, and V in J0931+0038 are quite low, which the authors correlate to other metal-poor stars in the bulge, halo, and dwarf galaxies. Fourth, they link the enhancement of Mn, Ni, and Zn to the stars discussed above. And finally/fifth, the first peak neutron capture elements from Sr and Pd are also enhanced, but [Ba/Fe] is quite low in concordance with some of the most extreme stars in ultra-faint dwarf galaxies.

\cite{ji2024spectacular} further constrain possible models by imposing a theoretical upper limit on the maximum achievable [Fe/H] after the dispersal of stellar yields and its dilution into the interstellar medium (ISM). For this, they first set a minimum diluting gas mass in their eq. 1, which is a function of explosion energy and ISM density, assuming spherical symmetry and a homogenous ISM. They then derive the upper limit on [Fe/H] in eq. 2, which is a function of supernova explosion energy, ISM density, and the mass of Fe ejected by the dying star. Using this upper limit and various models, they claim that such a high metallicity can only be achieved by stars more massive than 50 \(M_{\odot}\) or by PISN with initial mass $>$ 200 \(M_{\odot}\).

\cite{ji2024spectacular} limited their fit (that yielded the $80 M_{\odot} $ hypernova) with models from \cite{10.1093/mnras/sty1417} to only a subset of observations asserting that the full set of abundances cannot be fit by any existing model. Here, we list the full set of observations from Table 1 in \cite{ji2024spectacular} in our Table \ref{tab:2}. This set has yet another solar abundance standard, using the one from \cite{magg2022observational} and filling gaps therein from \cite{asplund2009chemical}. Again, for consistency, we apply the same standard in our analysis.

\begin{table}
    \centering
    \caption{Elemental abundances of J0931+0038 \citep{ji2024spectacular}.}
    \begin{threeparttable}
    \begin{tabular}{lcccccc}
        \hline
        \hline
        Species & N & $\mathrm{log}\epsilon$ & [X/H] & [X/Fe] & $\sigma$ & $\Delta_{NLTE}$ \\
        \hline
        Li I & 1 & 1.15 & 0.10 & 1.86 & 0.12 & $\cdots$\\
        C-H & 2 & 6.07 & -2.49 & -0.73 & 0.22 & 0.10\\
        N-H & 1 & 5.36 & -2.62 & -0.86 & limit & -0.10\\
        O I & 3 & 7.23 & -1.54 & 0.22 & 0.17 & -0.06\\
        Na I & 2 & 2.80 & -3.49 & -1.73 & 0.09 & -0.10\\
        Mg I & 8 & 5.52 & -2.03 & -0.27 & 0.11 & -0.02\\
        Al I & 1 & 4.15 & -2.28 & -0.52 & 0.50 & 0.65\\
        Si I & 8 & 5.80 & -1.79 & -0.03 & 0.13 & -0.17\\
        K I & 2 & 2.05 & -3.09 & -1.33 & 0.14 & -0.20\\
        Ca I & 30 & 4.73 & -1.64 & 0.12 & 0.13 & 0.02\\
        Sc II & 3 & -0.06 & -3.13 & -1.37 & 0.09 & $\cdots$\\
        Ti II & 30 & 2.56 & -2.38 & -0.62 & 0.10 & 0.12\\
        V II & 7 & 1.48 & -2.41 & -0.65 & 0.10 & $\cdots$\\
        Cr II & 5 & 3.75 & -1.99 & -0.23 & 0.12 & $\cdots$\\
        Mn I & 10 & 4.06 & -1.46 & 0.30 & 0.06 & 0.34\\
        Fe I & 181 & 5.74 & -1.76 & 0.00 & 0.13 & 0.11\\
        Co I & 4 & 3.37 & -1.58 & 0.18 & 0.10 & 0.11\\
        Ni I & 26 & 5.04 & -1.20 & 0.56 & 0.15 & 0.31\\
        Cu I & 1 & 2.64 & -1.55 & 0.21 & 0.19 & 0.35\\
        Zn I & 2 & 3.17 & -1.38 & 0.38 & 0.08 & $\cdots$\\
        Rb I & 1 & 2.61 & 0.09 & 1.85 & limit & $\cdots$\\
        Sr II & 3 & 1.85 & -1.02 & 0.74 & 0.07 & 0.03\\
        Y II & 24 & 1.27 & -0.94 & 0.82 & 0.11 & 0.00\\
        Zr II & 15 & 1.87 & -0.71 & 1.05 & 0.08 & $\cdots$\\
        Nb II & 1 & 1.34 & -0.12 & 1.64 & limit & $\cdots$\\
        Mo I & 1 & 0.55 & -1.33 & 0.43 & 0.16 & $\cdots$\\
        Ru I & 1 & 0.68 & -1.07 & 0.69 & 0.14 & $\cdots$\\
        Rh I & 1 & 0.34 & -0.57 & 1.19 & limit & $\cdots$\\
        Pd I & 1 & 0.46 & -1.11 & 0.65 & 0.14 & $\cdots$\\
        Ag I & 1 & -0.15 & -1.09 & 0.67 & limit & $\cdots$\\
        Ba II & 1 & -2.33 & -4.51 & -2.75 & 0.13 & 0.18\\
        Eu II & 1 & -1.79 & -2.31 & -0.55 & limit & $\cdots$\\
        \hline
    \end{tabular}
    
    \begin{tablenotes}
    \footnotesize
    \item In our notation: "$\sigma$ (uncertainty)" is $\sigma$ in \cite{ji2024spectacular}
    \end{tablenotes}
    
    \end{threeparttable}
    \label{tab:2}
\end{table}

\section{Underlying issues}
\label{issues}

We identify several problems with this picture that advocate against adopting their assumptions. Their tacit but central assumption is that either the entire supernova/hypernova yield is retained and mixed in the galaxy or that a representative fraction of yields is retained and lost, respectively. Even if we entertained the thought of a symmetric supernova explosion, a representative deposition of stellar yields into neighbouring clouds seems problematic, because the medium surrounding the explosion will be inhomogeneous with varying densities (e.g., collapsing clouds, filaments, etc.) such that ejecta with different radial positions in the explosion will likely be deposited in different fractions onto neighbouring clouds – some unknown degree of internal mixing will happen in the supernova shells and during the interaction with the surrounding ISM. This needs to be quantified in future models, but for now we cannot put a limit on this. Internal mixing during transit is limited by the $1000$ km/s expansion velocity, limiting transit time itself to $1$ kyr per parsec. In addition, the swept-up medium is likely to contribute a contamination, e.g., by stellar winds from other sources.

Yet, from both theory and observations, we know that the supernova explosion itself is non-symmetric (i.e., non-spherical). This has been known for decades, since the distribution of neutron stars (as the central remnants of supernova explosions) requires a large kick/momentum from the explosion, thus proving explosion/shell asymmetry \citep{10.1093/mnras/stw1275}. Further, this has now been shown more directly both in modelling of explosions \citep{vartanyan2019successful, muller2020hydrodynamics, van2023modelling, meyer2015asymmetric, vartanyan20253d} and observations \citep{Hughes_2000, Fesen_2006, Larsson_2013}. Various supernova explosions have been observed to have non-simplistic structures such as filaments and knots, such as in the Cassiopeia-A supernova remnant \citep{Hughes_2000, Fesen_2006} and supernova 1987A \citep{Larsson_2013, orlando2025tracing}. These asymmetries are driven by, for example, the Rayleigh-Taylor instability in neutrino-propelled supernovas, while convective and shock instabilities also play an essential role \citep{vartanyan2019successful, van2023modelling}. There can be other such mechanisms at play, like the lepton-number emission self-sustained asymmetry (LESA) \citep{Tamborra_2014} and the standing accretion shock instability (SASI) \citep{Blondin_2003}. Such instabilities can drive the supernova away from spherical symmetry very quickly. There are other flavours that we do not have the space to discuss here. Particularly interesting is that \cite{10.1093/mnras/sty1417} recognise that asymmetry in the supernova will modify the overall yields already towards patterns observed in metal-poor stars.

The bottom line is that we can expect the majority of SN explosions to be significantly asymmetrical, resulting in enrichment of neighbouring protostellar clouds directly with different fractions from different positions in the explosion. The more metal-poor a star, the more prominently these peculiar abundances show on the logarithmic scale.

\section{Theory and Methods}
\label{theory}
\subsection{Statistical model construction}

We can now cast the above discussion into a simplified model for stellar enrichment. We do not have the overall statistics for all metal-poor stars, but the freedom to select different weights for different shells (i.e., levels of mass coordinate/"radius" in their models) allows us to rephrase the fitting problem: each shell in the star has an abundance vector (i.e., its composition). Thus, we have to ask: is the observed abundance pattern within the vector space spanned by all these abundance vectors? I.e., we need to find a solution for:

\begin{equation}\label{eq1}
\sum_{i,j=1}^{n,m} a_i x_{ij} = y_{j}+ \epsilon_j; a_i \in (0,1)
\end{equation}

where \(i\) spans the different regions/shells the model has been divided into and \(j\) spans the different elements. $y$ is the observed abundance vector, and the $\epsilon_j$ is an error term, which budgets both model and observational uncertainty. $x_{i}$ is the abundance vector of each shell $i$. The model is solved for the coefficients \(a_i\), which are the relative contributions of each shell.

A few arguments force a simplification of the above solution: in principle there are hundreds of chemically different layers in a supernova model, and a full solution is computationally unfeasible. The differences between neighbouring $a_i$ should be constrainable from observations and theory. However, we are not aware of any current quantified constraints. This could be handled by some entropy condition/prior on the $a_i$ (limiting their differences). Overfitting risks are additionally reduced by lowering the number of fitted coefficients/shells in the star and checking after the fit that the distribution of coefficients is reasonable. We also note that the stellar structure is not just one of "burning shells" but is affected by a complex pattern of convection zones and added radial mixing (stronger in rotating stars). From star to star there are different main regions within which the abundances vary less steeply. 
We settled for only $7-8$ shells in total, a comparably low number that was sufficient to reflect the main abundance vector direction and still provide a good fit.

\subsubsection{Models}

For the yields $x_{ij}$ we use \cite{limongi2018}'s "normal" supernova models. For \(x\)'s and \(y\)'s we use the ratio of the number of atoms of the elements (\(N_X\)) to the number of hydrogen ($H$) atoms (\(N_H\)), i.e., \(N_X/N_H\), and we dilute the model ($a_i x_{ij}$ in $10 M_{\odot}$ of $H$ (this mass is arbitrary/largely irrelevant; its choice within reasonable limits corresponds to a simple scaling of the $a_i$). The resulting $a_i$ can be interpreted as the fraction of the supernova material that has polluted a pristine cloud of this size, and evidently we demand $a_i \in (0,1)$.

To compare the goodness of fits, we employ the same $\chi^2$ statistics as in eq. (3) from \cite{Skúladóttir_2024}:

\begin{equation}
    \chi^2 = \sum_{i=1}^{n} \frac{\left( [X_i/Fe]_{\rm obs} - [X_i/Fe]_{\rm model} \right)^2}{(\sigma^2_{[X_i/Fe]_{\rm obs}} + \sigma^2_{[X_i/Fe]_{\rm model}})}
    \label{eq:sum_over_xi_fe}
\end{equation}

where \([X_i/Fe]_{\rm obs,model}\) and \(\sigma_{[X_i/Fe]_{\rm obs,model}}\) are the observed and theoretical elemental abundance ratios and standard deviations, respectively. \cite{Skúladóttir_2024} assume \(\sigma^2_{[X_i/Fe]_{model}}\) = 0.25, and we adopt this. We note that this assumes the mutual independence (no assumed statistical correlation) of all theoretical and observational uncertainties, which should be refined in future work.

We fitted the observations tabulated in Tables \ref{tab:tab_1}, \ref{tab:1}, and \ref{tab:2} with various models from \cite{limongi2018}. For AS0039, we found the best fit case for the 13 \(\textup{M}_\odot\) model with an initial surface rotation velocity of 300 km/s and initial metallicity of [Fe/H] = -3. We compare the fit from \cite{Skúladóttir_2024} and ours in Fig. \ref{fig:main_fit} and discuss each element in the following subsections. We discuss only AS0039 in such detail, while the fits for HE 1327-2326 and J0931+0038 have been discussed succinctly.

\subsection{AS0039 fit}

\subsubsection{C abundance}
The estimated [C/Fe] value is very close to the observations in both our approach and \cite{Skúladóttir_2024}, with $-0.60$ dex and $-0.51$ dex for our and their case, respectively, i.e., with a residual of $0.01$ dex in our fit and $-0.08$ dex in their fit. They stress that the low [C/Fe] abundance of the star can only be explained with hypernova enrichment; we note that our model comfortably fits this.

\subsubsection{Na and Al abundances}
Na and Al are often grouped together due to both of them being light and odd atomic number nuclei. It is evident from Fig. \ref{fig:main_fit} that our fitting explains both of these elements better than \cite{Skúladóttir_2024}. Al is not only better but is almost a perfect fit. We note that higher Na/Al yield has been linked to mixing in rotating massive stars \citep{LANGER1997457}, and we use a rotating model. However, we would caution against solely putting this down to rotation, as in our fitting regime the mix of different shells becomes important.

\subsubsection{Mg, Si, Ca, and Ti abundances}

In general, alpha burning is rather standard, so we do not expect major issues in the ratios of alpha elements (with the exception of Ti, which could arguably be removed from the historical classification as an alpha element, as its chains are more complex and not within the main chain \citep{RevModPhys.29.547}). Consistently, both our fit and the hypernova models show similar behaviour. Our [Si/Fe] ratio is slightly higher and [Ti/Fe] slightly lower; we see a nearly perfect fit for Ca and a better agreement between observations and our fitting for Mg. It can be argued that increasing the explosion energy might help better our fitting for Ti, and hence these observations could be better explained by a higher energy supernova, but studies performed by \cite{10.1093/mnras/sty1417} show that explosion energy has minimal effect on the [Ti/Fe] ratio irrespective of the stellar mass. The same is reiterated by \cite{Skúladóttir_2024} using inferences from \cite{Heger2002}. Explosion energy would also not affect the [Si/Fe] ratio much \citep{10.1093/mnras/sty1417}. Principally, one could try to bring down [Si/Fe] by using different mass cuts and/or coefficients, but due to its irrelevance for the question of hypernova or normal supernova, we did not engage with it.

\subsubsection{Sc, Cr, Mn, Co, Ni, and Zn abundances}

The fitting of the iron peak elements is either better than or nearly the same as in the hypernova model. The observed [Ni/Fe] is in the middle of \cite{Skúladóttir_2024}'s and our fit. Interestingly, increasing the explosion energy would not help to bring the model ratio down \citep{10.1093/mnras/sty1417}, so the culprit is likely other model specifics (e.g., timescales and the enhancement with neutrons pre-explosion), as different codes are known to produce different yields for the same stellar models \citep{Farmer_2023}. [Zn/Fe] is not fit as well as other elements either. It appears to trend up with explosion energy at the $0.5 {\rm dex}$ level \citep{10.1093/mnras/sty1417}, not enough to capture the difference. We put little weight on this, since the $15 M_{\odot}$ model fits perfectly well on [Zn/Fe] (at the expense of [C/Fe] fitting worse). Compared to the hypernova model, our [Co/Fe] shows a similar fit. Generally, Sc is considered a central element for diagnosing a high explosion energy \citep{10.1093/mnras/sty1417}, and we have a nearly perfect fit for it. We have better [Cr/Fe] and [Mn/Fe] ratios. Mn is a perfect fit using our approach, in stark contrast to \cite{Skúladóttir_2024}'s fit. Interestingly, in \cite{10.1093/mnras/sty1417}, [Mn/Fe] is strongly anti-correlated to increasing explosion energy. \cite{Skúladóttir_2024} underpredict [Mn/Fe] with their hypernova model, so this mismatch is evidence against the hypernova picture, while our normal supernova predicts [Mn/Fe] correctly.

\begin{figure}
    \centering
    \includegraphics[width=1\linewidth]{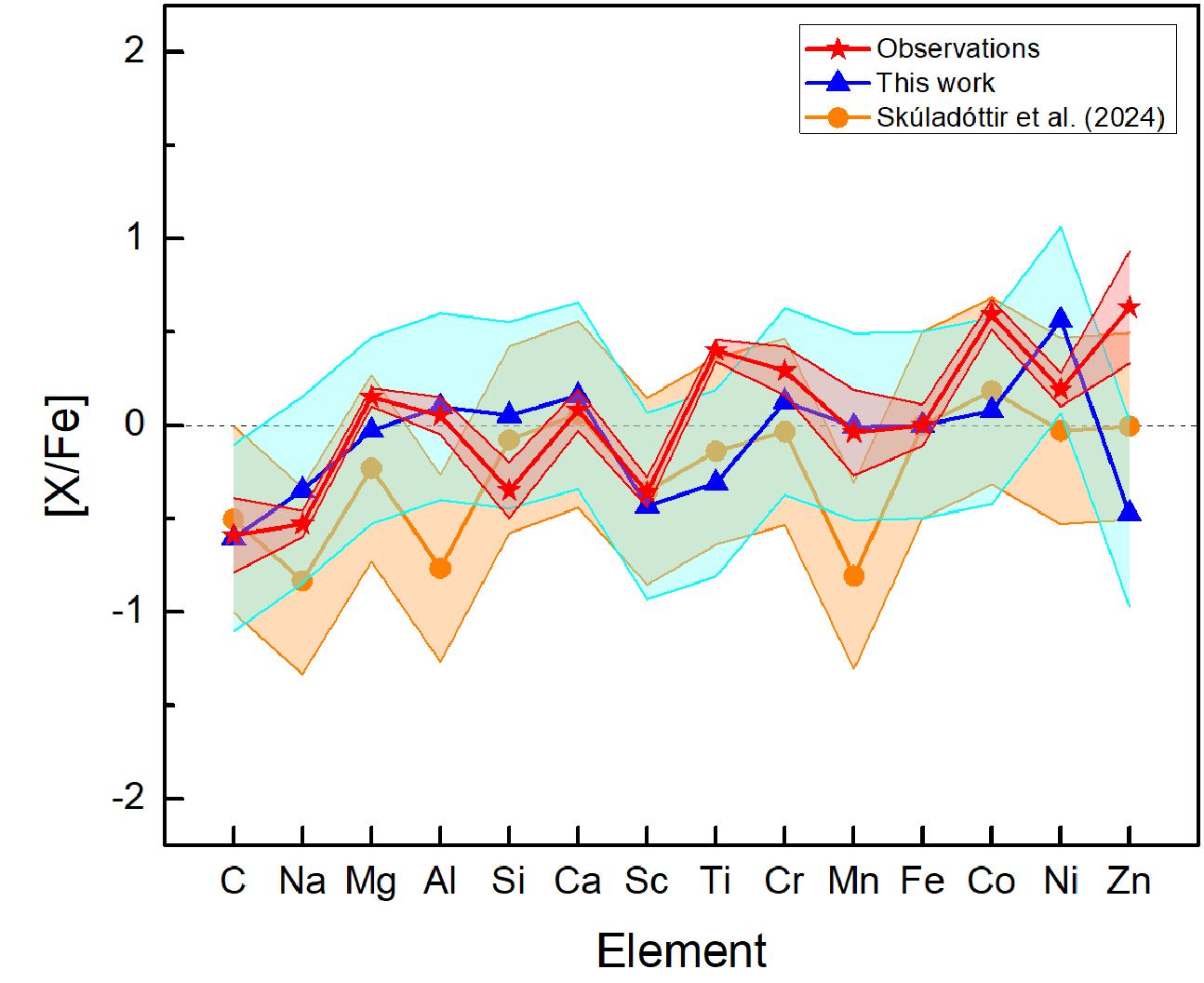}
    \caption{Comparison of \protect\cite{Skúladóttir_2024}'s 20 \(\textup{M}_\odot\) hypernova model fit and the predicted chemical abundance pattern obtained by using the method presented in this work with \protect\cite{limongi2018}'s 13 \(\textup{M}_\odot\) model.}
    \label{fig:main_fit}
\end{figure}

\begin{figure}
    \centering
    \includegraphics[width=1\linewidth]{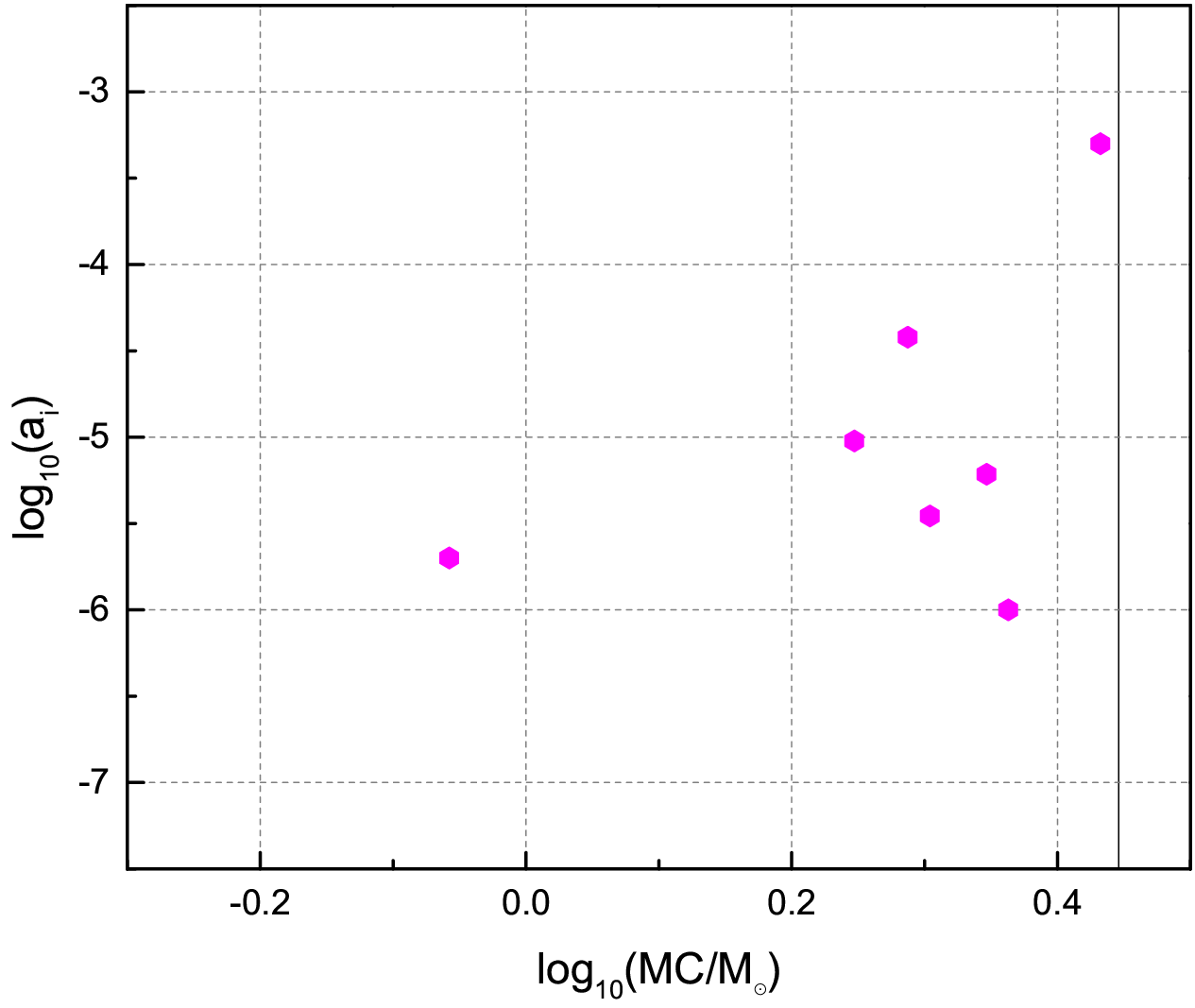}
    \caption{The log mass coordinate (MC) of the 13 \(\textup{M}_\odot\) model star vs. the contribution coefficients (\(a_i\), on a log scale) resulting from fitting the AS0039 observations. The values of coefficients have been marked at the beginning of their respective regions. The solid line marks the end of the last region.}
    \label{fig:coeff}
\end{figure}

\subsection{HE 1327-2326 fit}

Just like AS0039, we note here that the authors did – despite positing an asymmetric explosion – uniformly distribute the total yields. When we relinquish this assumption, the star is much more favourably explained by just a standard supernova explosion.

\begin{figure}
    \centering
    \includegraphics[width=1\linewidth]{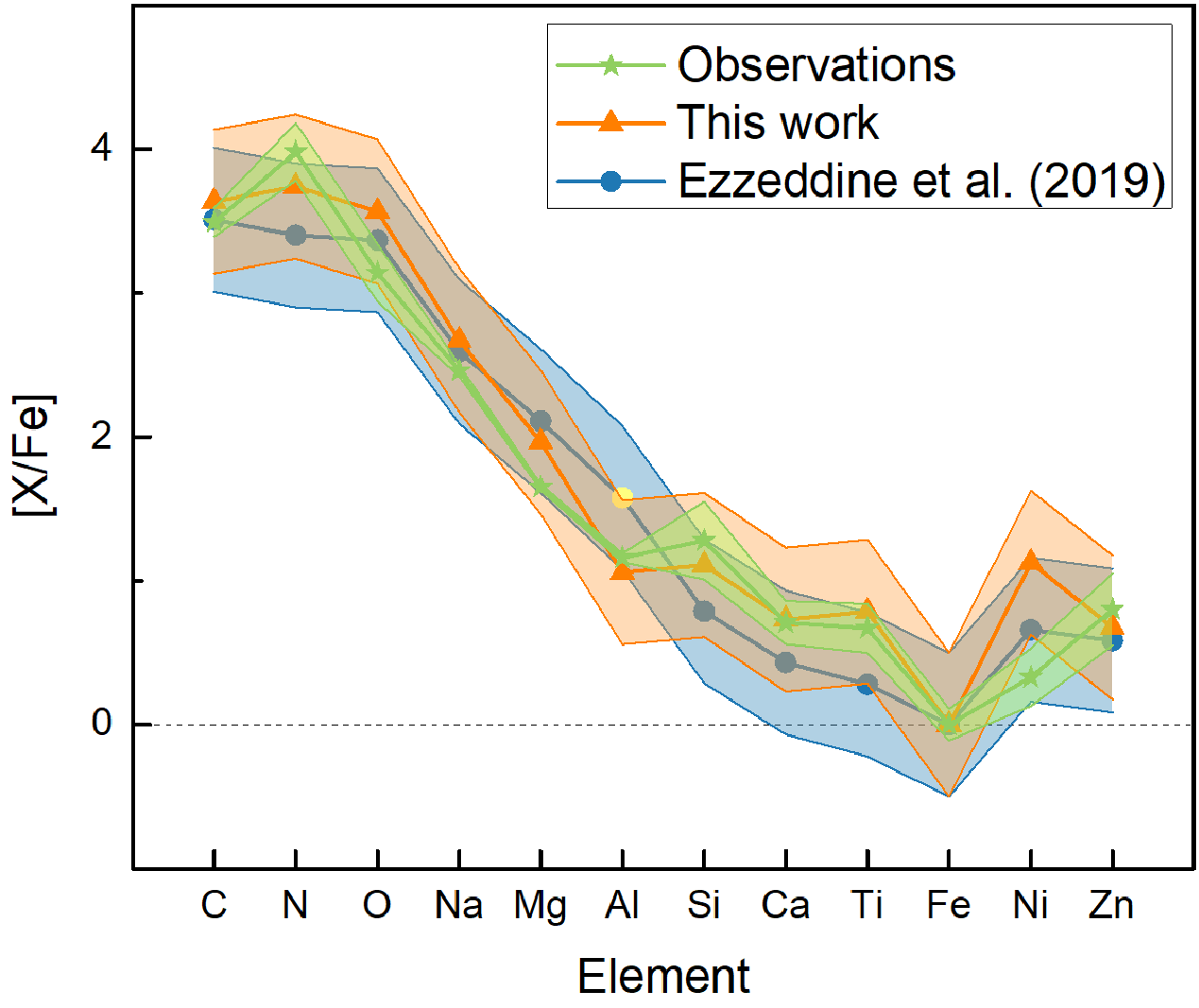}
    \caption{Comparison of our normal (15 \(M_{\odot}\) SN fit with the HE 1327-2326 hypernova fit (upper limits not shown). See Fig. \ref{fig:HE_ul_fit} for a plot with upper limits.}
    \label{fig:HE_main_fit}
\end{figure}

\begin{figure}
    \centering
    \includegraphics[width=1\linewidth]{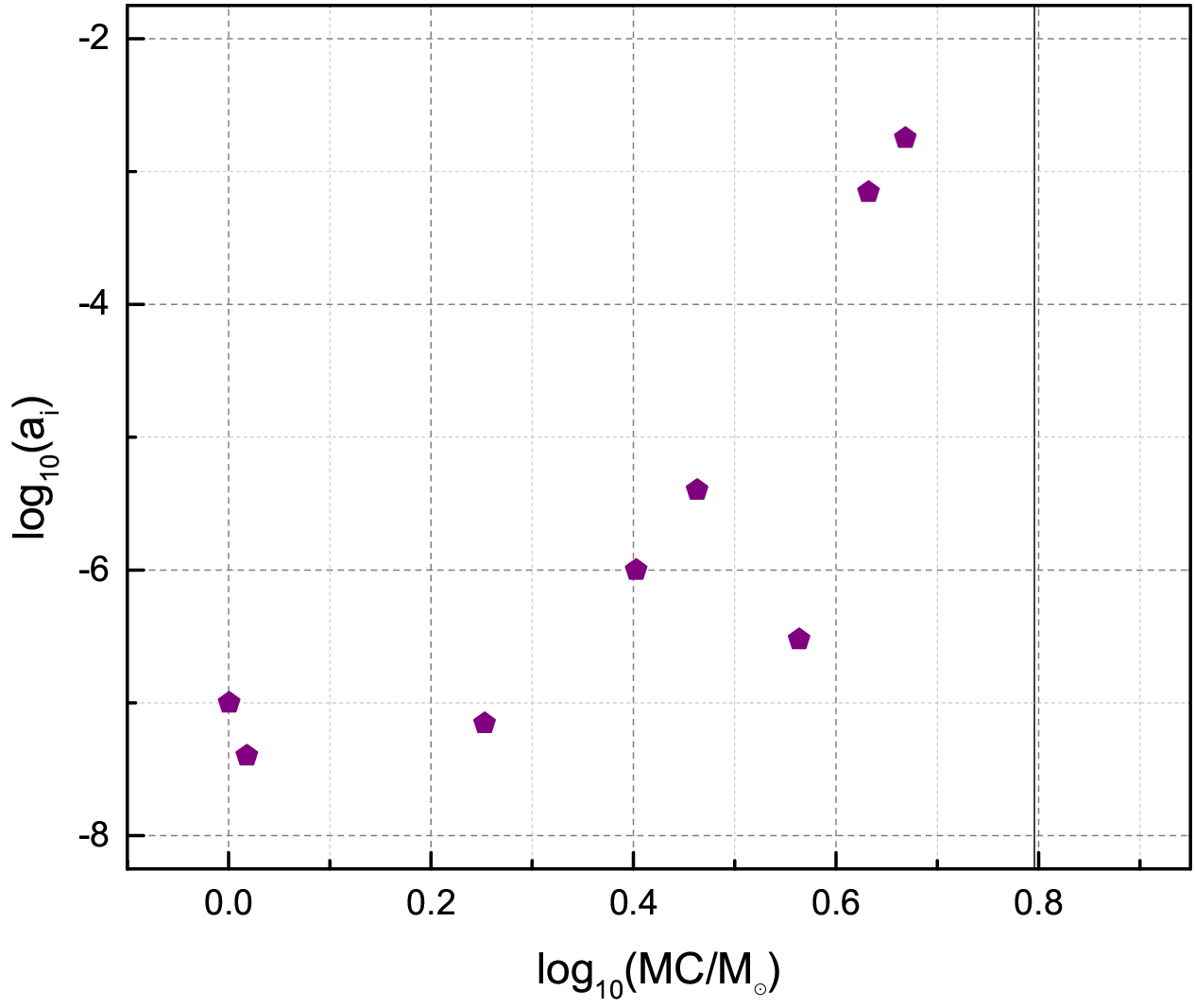}
    \caption{Same as Fig. \ref{fig:coeff}, but for 15 \(\textup{M}_\odot\) model star and HE 1327-2326.}
    \label{fig:HE_main_coeff}
\end{figure}

We compare our fit and the hypernova fit in Fig. \ref{fig:HE_main_fit}. \cite{ezzeddine2019evidence} do not fit Sr, and to maintain consistency, we do not either. We have used a 15 \(M_{\odot}\) supernova model that has a rotation speed of 300 km/s and has been initialised with a metallicity of [Fe/H] = -3 from \cite{limongi2018}. The different regions that the stellar model has been split into and the respective coefficients used have been presented in \ref{fig:HE_main_coeff}.

In Figs. \ref{fig:HE_main_fit} and \ref{fig:HE_main_coeff} we present our fit as for AS0039. Again, our coefficients (Fig. \ref{fig:HE_main_coeff}) form a consistent pattern with a simple trend overlaid by some scatter. Our fit performs better than the hypernova model for several elements, in particular also for Ti and Zn (originally taken as main arguments for a hypernova). The $\chi^{2}$ for our fit is $\sim$ 3.88 compared to theirs $\sim$ 5.05 (calculated as for AS0039). We note here that due to complexities in calculating the abundances of the hypernova model directly, we picked up the values from Fig. 4 of \cite{ezzeddine2019evidence}, and hence the values reported in our Fig. \ref{fig:HE_main_fit} are very close approximations of the true values. Assessment of the significance depends on one's prior about the model abundances (e.g., if one used $0.5$ dex as many authors do), which would lower the significance level. Although the use of a 0.5 dex error for the model values absorbs these approximation errors.

It is evident from the comparison of our and the hypernova fit that we do explain the stellar abundances of HE 1327-2326 better using just a simple supernova model, taking into account the procedure and arguments used for AS0039. Here as well, we note that the current reported fit is a good enough fit and certainly not the best. It can still be improved further using different models and/or shells.

\subsection{J0931+0038 fit}

\begin{figure}
    \centering
    \includegraphics[width=1\linewidth]{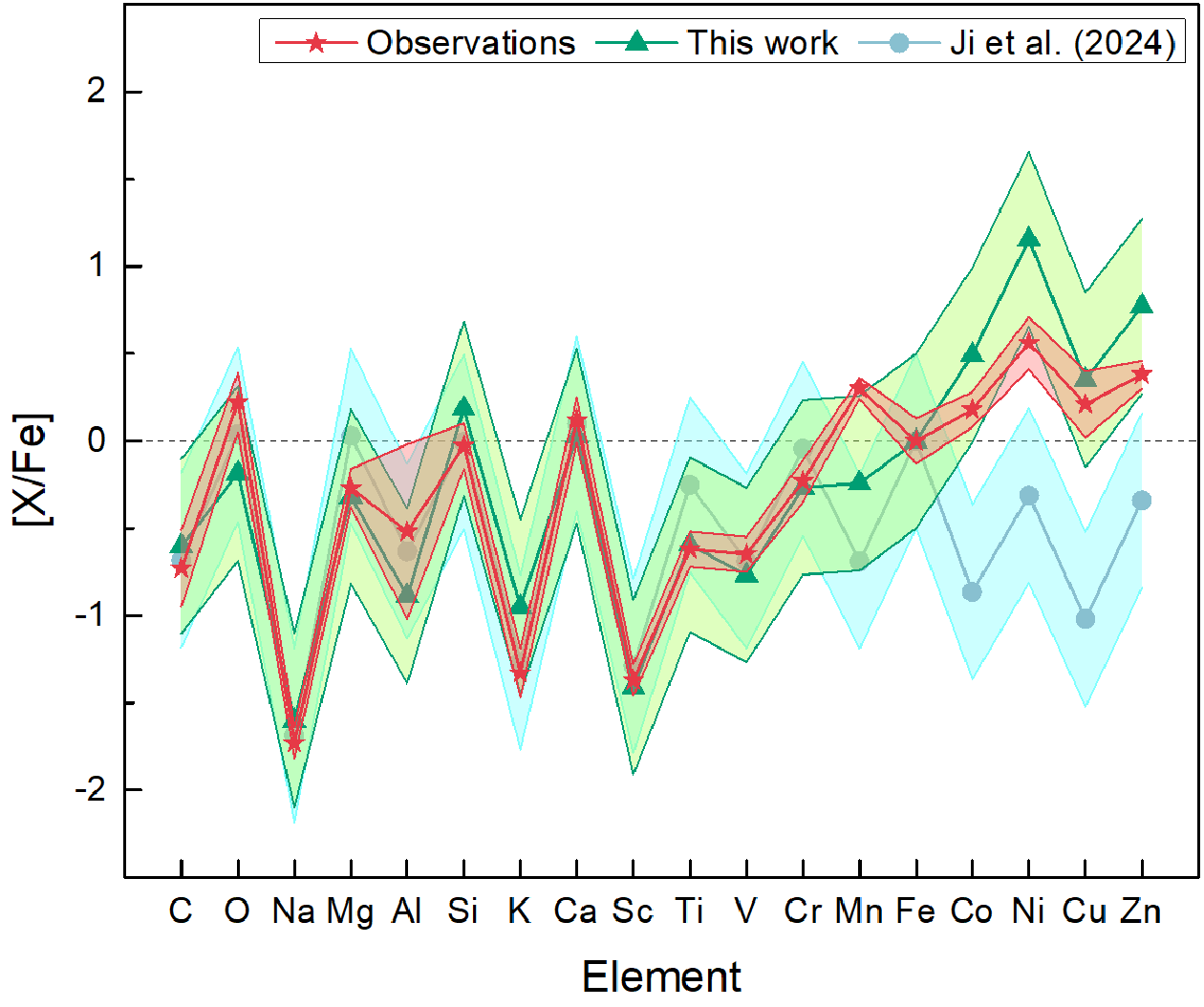}
    \caption{Comparison of our normal (13 \(M_{\odot}\) SN fit with the J0931+0038 hypernova fit (upper limits not shown). See Fig. \ref{fig:80_SM_ul_fit} for a plot with upper limits.}
    \label{fig:80_SM_main_fit}
\end{figure}

\begin{figure}
    \centering
    \includegraphics[width=1\linewidth]{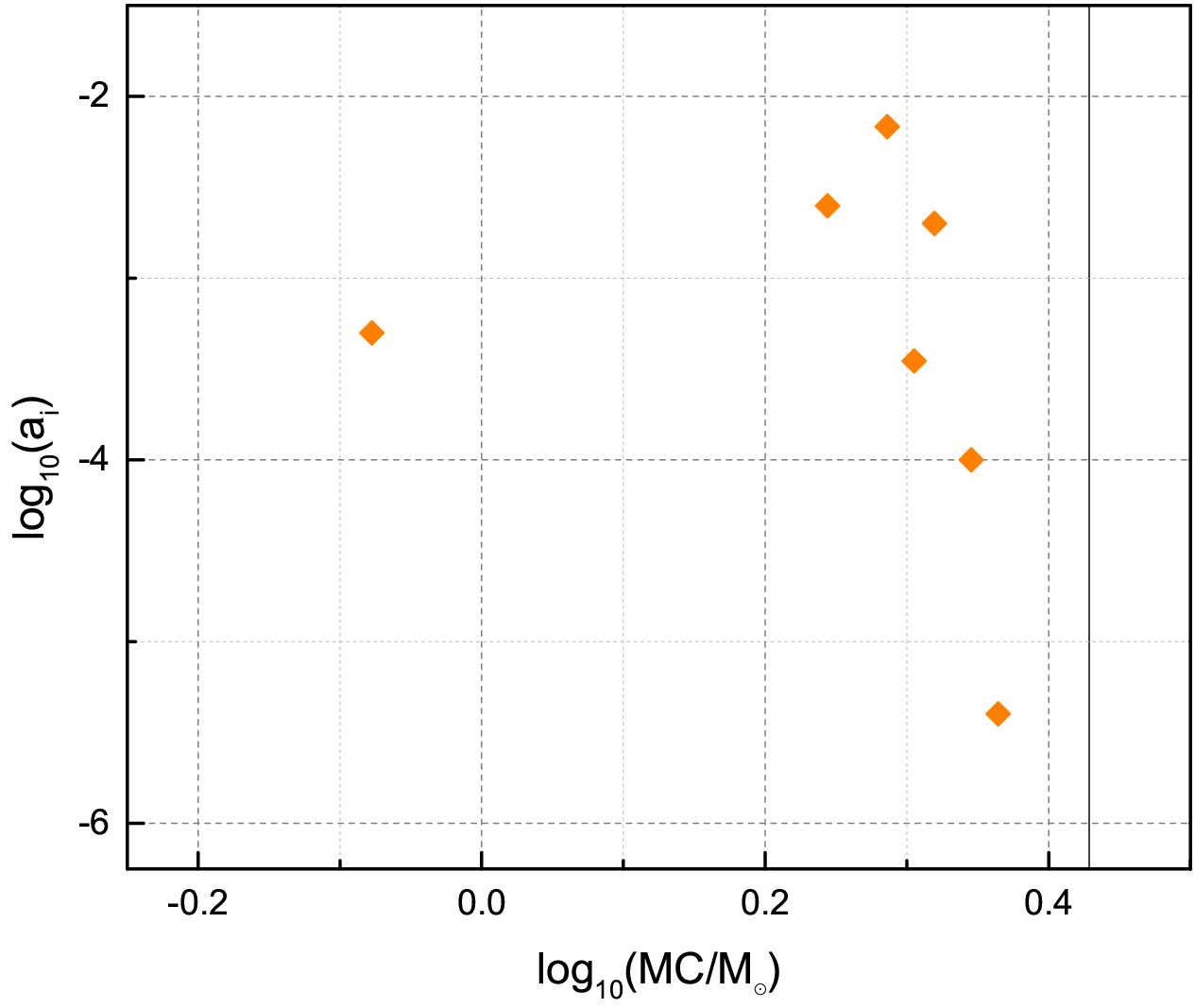}
    \caption{Same as Fig. \ref{fig:coeff}, but for a 13 \(\textup{M}_\odot\) model star and J0931+0038.}
    \label{fig:80_SM_main_coeff}
\end{figure}

We fit our model to the same subset of observations using the same fitting prescription and statistics as above, again with 0.5 dex uncertainty for the respective theoretical models. Here, we use a standard non-rotating 13 \(M_{\odot}\) model with initial metallicity ([Fe/H]) = -2 from \cite{limongi2018}. The comparison of our fit and the original work is presented in Fig. \ref{fig:80_SM_main_fit} with the corresponding coefficients/shells in Fig. \ref{fig:80_SM_main_coeff}.

Fig. \ref{fig:80_SM_main_fit} shows the clear superiority of our fit, which has $\chi^2 \sim 5.30$ versus $\sim 19.37$ for the hypernova. Indeed, \cite{ji2024spectacular} report their difficulty in fitting the Fe peak. Their model could not simultaneously produce low [Sc, Ti, V/Fe], which they link to low explosion energy; high [Ni, Zn/Fe], which they link to high explosion energy; and high [Mn/Fe], which they link to high neutron fractions while producing a small odd-even effect, which they link to low neutron fractions. Our fit – separating the SN into seven regions – alleviates these tensions.

However, we do note here that (following the original analysis) the fit is only for a subset of observations, and we have not yet exhausted a search among the available supernova models that would likely further improve the result. We leave this for future analysis, as this quick result already makes our point that the asymmetric dispersal of yields from different regions of a dying star must be taken into account and delivers better results than extraordinary models with too simplistic assumptions about yield dispersal.

\section{Results and discussion}
\label{res}

In Table \ref{tab:all_stats} we list our standardised chi-squared values for each of the literature hypernova claims, and compare them to the results from our analysis. 
The elemental abundance patterns are compared in Figs. \ref{fig:main_fit}, \ref{fig:HE_main_fit}, and \ref{fig:80_SM_main_fit} for AS0039, HE 1327-2326, and J0931+0038, respectively.

Three other fits to the stars have been presented in Figs. \ref{fig:15_SM}, \ref{fig:HE_ul_fit}, and \ref{fig:80_SM_ul_fit} using either the same stellar model or a model with a mass similar to that for the main fits. All of them yield a superior $\chi^2$ compared to the ones obtained from original hypernova fits.

\begin{table}
    \centering
    \begin{tabular}{lcc}
    \hline
    \hline
        Star & Stand. ${\chi^2}_{(Literature)}$ & Stand. ${\chi^2}_{(This work)}$\\
        \hline
        AS0039 & 9.33 & 8.14\\
        HE 1327-2326 & 5.05 & 3.88\\
        J0931+0038 & 19.37 & 5.30\\
    \hline
    \end{tabular}
    \caption{Comparison of standardised ${\chi^2}$ values of all stars.}
    \label{tab:all_stats}
\end{table}

We only use the interior shells (i.e., within the helium-burning region), as we wanted to stay clear of using a region of pre-set solar-scaled abundances found in the outermost shell. In addition, a major part of the outer shell is likely lost in the wind before the supernova explosion and may or may not end up in the target protostellar cloud. 

The patterns of coefficients are found in Figs. \ref{fig:coeff}, \ref{fig:HE_main_coeff}, and \ref{fig:80_SM_main_coeff} respectively, for each star. As we have no reliable limits on autocorrelation between neighbouring shells (these will have to be derived from both 3D explosion models and also full models for later redistribution of gas and the evolution of the ejecta), we did not impose ourselves any constraints on correlation between the coefficients of neighbouring shells. Despite this (and also despite the uncertainties in both model and abundance data), the data themselves enforce a significant autocorrelation between neighbouring shells. Common trends (either up or down with shell location) are visible to the naked eye and the DW test statistic for the overall sample is at $1.14$ vs. a $5 \%$ significance level at $1.24$, i.e. the autocorrelation is highly significant (see Fig. \ref{fig:all_coeff}). This suggests that i) these results have a meaningful physical interpretation and ii) in principle this type of analysis can in the future on larger and more precise samples provide a direct measure of the autocorrelations.

We could be using a parametric fit over all radii in the star (i.e., using more, thinner zones, paying with computational cost). Although, for now, we have the freedom to choose an arbitrary number of layers with some constraints. Figs. 19 and 20 in \cite{Wang_and_Burrows_2024} show a simulation in support of our approach where different layers of the asymmetric supernova are expelled in different directions, resulting in large composition differences across the explosion within the first seconds, with an unknown final state.

Regarding our choice of shells: Ultimately a large number of shells constrained by some entropy condition might be better but poses a computational challenge.

Here we chose to instead use a minimal number of fitted shells. These shells were chosen to represent main regions of elemental composition, which are not equivalent to burning shells and also still have significant composition changes with radius. Consequently, our reported chi-squares are upper bounds to these models but sufficient to make our point that already the fits are better than the benchmarks.

To reiterate: we are trying to incorporate in our methodology the non-uniform dispersal of yields in the surroundings of the supernova. The main business (for future papers) will be quantifying the inhomogeneity of the supernova explosion itself and quantifying the mixing post-explosion. However, this is a major endeavour far beyond the work of this single paper that is trying to start this initiative.

\section{Observability in modern stars}
\label{obs_ques}

While we discuss metal-poor peculiar stars in this paper, the connections to observations at higher metallicity and in the present-day Milky Way need to be drawn.

The main complication for this is the high baseline metallicity (the metals present in the star-forming region pre-supernova enrichment) at the present day. To formalise this thought: think of a model where you have a (fixed, metallicity-independent) distribution of peculiar yields $f_p(x_i)$ in element i (on an absolute mass scale) contributed by stochastic processes plus a baseline abundance $b_i$. The resulting distribution of elements reads:
\begin{equation}
\label{baseline_eqn}
f(y_i) = b_i + f_p(x_i)
\end{equation}
On the logarithmic scale of relative abundances, the baseline $b_i$ then dictates the scale of the observable scatter – approximately an order of magnitude per dex in baseline metallicity; e.g., a scatter of $0.3$ dex at a baseline metallicity of $-1.5$ dex would correspond to a scatter of $0.01$ dex at solar metallicity.

Indeed, present-day abundances in local stars have been found to be very homogeneous; see, e.g., \cite{nieva2012present}. Solar twins have enabled more precise relative analysis, revealing scatter at the $\sim 0.01$ dex level due to the larger baseline metallicity \citep{Nissen1, Nissen2, Nissen3}, instead of the $0.5$ dex for very metal-poor stars, where the principal trends have been associated with condensation temperature and planet formation \citep{melendez2009, Sun2025}.

To demonstrate this effect, we transposed the problem to solar metallicity, i.e., we assume mixing with a solar-composition cloud for the observations. The solar composition is set such that the metallicity ([Fe/H]) of the final mixture equates to zero. I.e., $b_i$ in eqn. \ref{baseline_eqn} is now non-zero for all the elements. This reduced the deviation in element ratios from the baseline on the [X/Fe] scale accordingly. The largest transposed deviations are $\sim 3.25 \times 10^{-4}$ for Zn in AS0039, 0.008 dex (N, HE 1327-2326), and $\sim$ 0.02 dex (Ni, J0931+0038), respectively, for the solar-scaled case. So, on an absolute scale, the most metal-rich object, J0931+0038, would, ceteris paribus, still display detectable peculiarities.

Thus, the equivalent scatter in low-metallicity stars could be detectable in solar twins, and it will be an interesting task for future work to build a quantitative model for stochasticity at low metallicity and compare it to scatter in solar twins. Another target will be short- and medium-lifetime radioactive elements, where no significant baseline effect exists. These elements are expected to have been quite directly delivered by nearby nucleosynthesis in the solar system \citep{CHAUSSIDON2007872, huss2009stellar, davis2022short}. While there are few direct observations (e.g., on $^{26}{\rm Al}$), there is a more detailed source of information: short-lived isotopes in the protosolar nebula that are imprinted on meteorite abundances; see e.g., \citep{wasserburg2006short}.

A priori, we can see that such a quantitative stochastic enrichment model for the whole population will have an interesting, testable prediction: different groups of elements are expected to show a different size of variation/scatter between stars \citep{cayrel2004first, Francois2007, Hansen2012, roederer2014_light, roederer2014_heavy} (more scatter for elements with lower baseline abundance and produced in specific/unique regions), which depends on their production fractions in the models and the SN instabilities at play. As above, we expect this effect of different scatter by element group to be measurable across metallicities and to – after disentangling from other sources of peculiarity, like the larger Poisson noise expected for r-process elements or condensation temperature dependencies linked to planet formation \citep{melendez2009} – help quantify the mixing stochasticity.

\section{Conclusions}
\label{discuss}

The main message from this study is that more care needs to be taken when inferring particular explosion mechanisms from single low-metallicity stars with peculiar abundances: while the field uses the widespread assumption that one metal-poor star is enriched by the mix of yields from a single super/hypernova, we advocate instead for enrichment by fractions of a supernova. In reality, not one entire supernova remnant/bubble gets absorbed into a nearby star-forming cloud, but pieces of an inherently asymmetric explosion. 
For the stars in question, we found – contrary to the previous assessments taking them as proof of hypernovae – that instead these stars are better matched by a standard supernova model.

We note the plethora of caveats to be made in this study. As noted in \cite{Skúladóttir_2024}, stellar evolution models and reaction chains are highly uncertain, reflected by $0.5$ ${\rm dex}$ model uncertainty. For AS0039, we thus made sure to focus our discussion and fits on elements which are claimed to show a clear dependence on explosion energy/entropy (C, Sc, Mn, Zn). Consequently, we show a model that performs well on these elements but point out that other models principally yield equally good fits in general. In Fig. \ref{fig:15_SM} we show an alternative progenitor mass (15 \(\textup{M}_\odot\)) which has a $\chi^2 \sim 7.70$ with only $4$ shells used. This model has the same rotation speed and initial metallicity as that of the 13 \(\textup{M}_\odot\) model.

This also opens another argument on our approach. In principle, the formal question is if a star's abundance vector is contained in the span of all abundance vectors found in each model. However, the dimensionality is too high to attempt a formal fit. Since we could easily obtain better fits than our benchmarks, we also refrained from attempting more resource-consuming fits, which may become necessary in future investigations. To name an example: The elements Co, Ni, and Zn come from a similar central region, but within our shell limits, there is significant variation in their abundances, which in principle should allow for a better result for AS0039. However, these patterns also greatly depend on the rotation/mixing behaviour in the progenitor core, which means that more research is needed to understand these complex models.

Of course, this case can be generalised to other investigations. Each object needs to be reassessed along these same lines, and the bar for proving special progenitors is now set higher, as it must be shown (within this new regime) that a standard model cannot cover the observed patterns.  We still expect that some reliable inference can be made, especially by adopting a more physically motivated approach to larger samples.

In this study we had to operate in empty space, because no reliable predictions exist how much internal mixing within the stellar explosions and then external mixing happens before the yields get incorporated into new forming stars in the surroundings. The explosion asymmetry and those interactions need to be predicted from models to be then compared e.g. with the autocorrelations from the coefficients we made. Also, to go beyond our qualitative opening, for a quantitative study, full attention needs to be paid to correlated bias in both the spectroscopic analysis and the nucleosynthesis.

In this paper, we have raised the need for and ability of mixing stochasticity to explain peculiar abundances of stars. Like in all fields of research, quantification should follow the proof of principle. Here, the quantification will be in i) predictions of the abundance scatter of different elements in whole populations of stars, and we have outlined observability questions in Section \ref{obs_ques}. And ii) in using the quantified scatter to constrain mixing in supernova models and between supernova model ejecta and the star-forming ISM to compare and challenge simulations.

\section*{Acknowledgements}

The authors thank Marco Limongi and Alessandro Chieffi for providing the explosive nucleosynthesis yields for their models used in this work. The authors also thank M. Bergemann for thorough discussions. R.S. acknowledges the generous funding of a Royal Society University Research Fellowship. A.A. acknowledges funding from the University College London Research Excellence Scholarship.

\section*{Data Availability}

The main statistics have been presented in the paper. We are happy to provide any additional information upon request.



\bibliographystyle{mnras}
\bibliography{example} 

\begin{thebibliography}{}
\makeatletter
\relax
\def\mn@urlcharsother{\let\do\@makeother \do\$\do\&\do\#\do\^\do\_\do\%\do\~}
\def\mn@doi{\begingroup\mn@urlcharsother \@ifnextchar [ {\mn@doi@} {\mn@doi@[]}}
\def\mn@doi@[#1]#2{\def\@tempa{#1}\ifx\@tempa\@empty \href {http://dx.doi.org/#2} {doi:#2}\else \href {http://dx.doi.org/#2} {#1}\fi \endgroup}
\def\mn@eprint#1#2{\mn@eprint@#1:#2::\@nil}
\def\mn@eprint@arXiv#1{\href {http://arxiv.org/abs/#1} {{\tt arXiv:#1}}}
\def\mn@eprint@dblp#1{\href {http://dblp.uni-trier.de/rec/bibtex/#1.xml} {dblp:#1}}
\def\mn@eprint@#1:#2:#3:#4\@nil{\def\@tempa {#1}\def\@tempb {#2}\def\@tempc {#3}\ifx \@tempc \@empty \let \@tempc \@tempb \let \@tempb \@tempa \fi \ifx \@tempb \@empty \def\@tempb {arXiv}\fi \@ifundefined {mn@eprint@\@tempb}{\@tempb:\@tempc}{\expandafter \expandafter \csname mn@eprint@\@tempb\endcsname \expandafter{\@tempc}}}

\bibitem[\protect\citeauthoryear{Asplund, Grevesse, Sauval  \& Scott}{Asplund et~al.}{2009}]{asplund2009chemical}
Asplund M.,  Grevesse N.,  Sauval A.~J.,   Scott P.,  2009, Annual review of astronomy and astrophysics, 47, 481

\bibitem[\protect\citeauthoryear{Asplund, Amarsi  \& Grevesse}{Asplund et~al.}{2021}]{asplund2021chemical}
Asplund M.,  Amarsi A.,   Grevesse N.,  2021, Astronomy \& Astrophysics, 653, A141

\bibitem[\protect\citeauthoryear{Blondin, Mezzacappa  \& DeMarino}{Blondin et~al.}{2003}]{Blondin_2003}
Blondin J.~M.,  Mezzacappa A.,   DeMarino C.,  2003, \mn@doi [The Astrophysical Journal] {10.1086/345812}, 584, 971

\bibitem[\protect\citeauthoryear{Bray \& Eldridge}{Bray \& Eldridge}{2016}]{10.1093/mnras/stw1275}
Bray J.~C.,  Eldridge J.~J.,  2016, \mn@doi [Monthly Notices of the Royal Astronomical Society] {10.1093/mnras/stw1275}, 461, 3747

\bibitem[\protect\citeauthoryear{Burbidge, Burbidge, Fowler  \& Hoyle}{Burbidge et~al.}{1957}]{RevModPhys.29.547}
Burbidge E.~M.,  Burbidge G.~R.,  Fowler W.~A.,   Hoyle F.,  1957, \mn@doi [Rev. Mod. Phys.] {10.1103/RevModPhys.29.547}, 29, 547

\bibitem[\protect\citeauthoryear{Cassinelli}{Cassinelli}{1979}]{cassinelli1979stellar}
Cassinelli J.~P.,  1979, In: Annual review of astronomy and astrophysics. Volume 17.(A79-54126 24-90) Palo Alto, Calif., Annual Reviews, Inc., 1979, p. 275-308., 17, 275

\bibitem[\protect\citeauthoryear{Cavallo, Cescutti  \& Matteucci}{Cavallo et~al.}{2021}]{10.1093/mnras/stab281}
Cavallo L.,  Cescutti G.,   Matteucci F.,  2021, \mn@doi [Monthly Notices of the Royal Astronomical Society] {10.1093/mnras/stab281}, 503, 1

\bibitem[\protect\citeauthoryear{Cayrel et~al.,}{Cayrel et~al.}{2004}]{cayrel2004first}
Cayrel R.,  et~al., 2004, Astronomy \& Astrophysics, 416, 1117

\bibitem[\protect\citeauthoryear{Chaussidon \& Gounelle}{Chaussidon \& Gounelle}{2007}]{CHAUSSIDON2007872}
Chaussidon M.,  Gounelle M.,  2007, \mn@doi [Comptes Rendus Geoscience] {https://doi.org/10.1016/j.crte.2007.09.005}, 339, 872

\bibitem[\protect\citeauthoryear{Colavitti, Matteucci  \& Murante}{Colavitti et~al.}{2008}]{colavitti2008}
Colavitti E.,  Matteucci F.,   Murante G.,  2008, \mn@doi [A&A] {10.1051/0004-6361:200809413}, 483, 401

\bibitem[\protect\citeauthoryear{Cooke \& Madau}{Cooke \& Madau}{2014}]{cooke2014carbon}
Cooke R.~J.,  Madau P.,  2014, The Astrophysical Journal, 791, 116

\bibitem[\protect\citeauthoryear{Davis}{Davis}{2022}]{davis2022short}
Davis A.~M.,  2022, Annual Review of Nuclear and Particle Science, 72, 339

\bibitem[\protect\citeauthoryear{Deutsch}{Deutsch}{1956}]{deutsch1956circumstellar}
Deutsch A.~J.,  1956, Astrophysical Journal, vol. 123, p. 210, 123, 210

\bibitem[\protect\citeauthoryear{Ezzeddine et~al.,}{Ezzeddine et~al.}{2019}]{ezzeddine2019evidence}
Ezzeddine R.,  et~al., 2019, The Astrophysical Journal, 876, 97

\bibitem[\protect\citeauthoryear{Farmer, Laplace, Ma, de Mink  \& Justham}{Farmer et~al.}{2023}]{Farmer_2023}
Farmer R.,  Laplace E.,  Ma J.-z.,  de Mink S.~E.,   Justham S.,  2023, \mn@doi [The Astrophysical Journal] {10.3847/1538-4357/acc315}, 948, 111

\bibitem[\protect\citeauthoryear{Fesen et~al.,}{Fesen et~al.}{2006}]{Fesen_2006}
Fesen R.~A.,  et~al., 2006, \mn@doi [The Astrophysical Journal] {10.1086/504254}, 645, 283

\bibitem[\protect\citeauthoryear{{Fran{\c{c}}ois} et~al.,}{{Fran{\c{c}}ois} et~al.}{2007}]{Francois2007}
{Fran{\c{c}}ois} P.,  et~al., 2007, \mn@doi [\aap] {10.1051/0004-6361:20077706}, \href {https://ui.adsabs.harvard.edu/abs/2007A&A...476..935F} {476, 935}

\bibitem[\protect\citeauthoryear{Grimmett, Heger, Karakas  \& Müller}{Grimmett et~al.}{2018}]{10.1093/mnras/sty1417}
Grimmett J.~J.,  Heger A.,  Karakas A.~I.,   Müller B.,  2018, \mn@doi [Monthly Notices of the Royal Astronomical Society] {10.1093/mnras/sty1417}, 479, 495

\bibitem[\protect\citeauthoryear{{Hansen} et~al.,}{{Hansen} et~al.}{2012}]{Hansen2012}
{Hansen} C.~J.,  et~al., 2012, \mn@doi [\aap] {10.1051/0004-6361/201118643}, \href {https://ui.adsabs.harvard.edu/abs/2012A&A...545A..31H} {545, A31}

\bibitem[\protect\citeauthoryear{{Hartwig}, {Bromm}, {Klessen}  \& {Glover}}{{Hartwig} et~al.}{2015}]{Hartwig2015}
{Hartwig} T.,  {Bromm} V.,  {Klessen} R.~S.,   {Glover} S. C.~O.,  2015, \mn@doi [\mnras] {10.1093/mnras/stu2740}, \href {https://ui.adsabs.harvard.edu/abs/2015MNRAS.447.3892H} {447, 3892}

\bibitem[\protect\citeauthoryear{Heger \& Woosley}{Heger \& Woosley}{2002}]{Heger2002}
Heger A.,  Woosley S.~E.,  2002, The Astrophysical Journal, 567, 532

\bibitem[\protect\citeauthoryear{Heger \& Woosley}{Heger \& Woosley}{2010}]{Heger_2010}
Heger A.,  Woosley S.~E.,  2010, The Astrophysical Journal, 724, 341

\bibitem[\protect\citeauthoryear{Hughes, Rakowski, Burrows  \& Slane}{Hughes et~al.}{1999}]{Hughes_2000}
Hughes J.~P.,  Rakowski C.~E.,  Burrows D.~N.,   Slane P.~O.,  1999, \mn@doi [The Astrophysical Journal] {10.1086/312438}, 528, L109

\bibitem[\protect\citeauthoryear{Huss, Meyer, Srinivasan, Goswami  \& Sahijpal}{Huss et~al.}{2009}]{huss2009stellar}
Huss G.~R.,  Meyer B.~S.,  Srinivasan G.,  Goswami J.~N.,   Sahijpal S.,  2009, Geochimica et Cosmochimica Acta, 73, 4922

\bibitem[\protect\citeauthoryear{{Hutter}, {Cueto}, {Dayal}, {Gottl{\"o}ber}, {Trebitsch}  \& {Yepes}}{{Hutter} et~al.}{2025}]{Hutter2025}
{Hutter} A.,  {Cueto} E.~R.,  {Dayal} P.,  {Gottl{\"o}ber} S.,  {Trebitsch} M.,   {Yepes} G.,  2025, \mn@doi [\aap] {10.1051/0004-6361/202452460}, \href {https://ui.adsabs.harvard.edu/abs/2025A&A...694A.254H} {694, A254}

\bibitem[\protect\citeauthoryear{Iwamoto et~al.,}{Iwamoto et~al.}{1998}]{iwamoto1998hypernova}
Iwamoto K.,  et~al., 1998, Nature, 395, 672

\bibitem[\protect\citeauthoryear{Jeena \& Banerjee}{Jeena \& Banerjee}{2024}]{Jeena2024Origin}
Jeena S.~K.,  Banerjee P.,  2024, \mn@doi [The Open Journal of Astrophysics] {10.33232/001c.124113}, 7

\bibitem[\protect\citeauthoryear{Jeena, Banerjee  \& Heger}{Jeena et~al.}{2024}]{jeena2024core}
Jeena S.,  Banerjee P.,   Heger A.,  2024, Monthly Notices of the Royal Astronomical Society, 527, 4790

\bibitem[\protect\citeauthoryear{Ji et~al.,}{Ji et~al.}{2024}]{ji2024spectacular}
Ji A.~P.,  et~al., 2024, The Astrophysical Journal Letters, 961, L41

\bibitem[\protect\citeauthoryear{Karlsson \& Gustafsson}{Karlsson \& Gustafsson}{2005}]{karlsson2005}
Karlsson T.,  Gustafsson B.,  2005, \mn@doi [A&A] {10.1051/0004-6361:20042168}, 436, 879

\bibitem[\protect\citeauthoryear{{Koch}, {McWilliam}, {Grebel}, {Zucker}  \& {Belokurov}}{{Koch} et~al.}{2008}]{Koch2008}
{Koch} A.,  {McWilliam} A.,  {Grebel} E.~K.,  {Zucker} D.~B.,   {Belokurov} V.,  2008, \mn@doi [\apjl] {10.1086/595001}, \href {https://ui.adsabs.harvard.edu/abs/2008ApJ...688L..13K} {688, L13}

\bibitem[\protect\citeauthoryear{Korn, Grundahl, Richard, Mashonkina, Barklem, Collet, Gustafsson  \& Piskunov}{Korn et~al.}{2007}]{Korn_2007}
Korn A.~J.,  Grundahl F.,  Richard O.,  Mashonkina L.,  Barklem P.~S.,  Collet R.,  Gustafsson B.,   Piskunov N.,  2007, \mn@doi [The Astrophysical Journal] {10.1086/523098}, 671, 402

\bibitem[\protect\citeauthoryear{Langer, Fliegner, Heger  \& Woosley}{Langer et~al.}{1997}]{LANGER1997457}
Langer N.,  Fliegner J.,  Heger A.,   Woosley S.,  1997, \mn@doi [Nuclear Physics A] {https://doi.org/10.1016/S0375-9474(97)00290-X}, 621, 457

\bibitem[\protect\citeauthoryear{Larsson et~al.,}{Larsson et~al.}{2013}]{Larsson_2013}
Larsson J.,  et~al., 2013, \mn@doi [The Astrophysical Journal] {10.1088/0004-637X/768/1/89}, 768, 89

\bibitem[\protect\citeauthoryear{{Li}, {Leja}, {Johnson}, {Tacchella}  \& {Naidu}}{{Li} et~al.}{2024}]{tacchella}
{Li} Y.,  {Leja} J.,  {Johnson} B.~D.,  {Tacchella} S.,   {Naidu} R.~P.,  2024, \mn@doi [\apjl] {10.3847/2041-8213/ad5280}, \href {https://ui.adsabs.harvard.edu/abs/2024ApJ...969L...5L} {969, L5}

\bibitem[\protect\citeauthoryear{Limongi \& Chieffi}{Limongi \& Chieffi}{2018}]{limongi2018}
Limongi M.,  Chieffi A.,  2018, The Astrophysical Journal Supplement Series, 237, 13

\bibitem[\protect\citeauthoryear{Magg et~al.,}{Magg et~al.}{2022}]{magg2022observational}
Magg E.,  et~al., 2022, Astronomy \& Astrophysics, 661, A140

\bibitem[\protect\citeauthoryear{{Mel{\'e}ndez}, {Asplund}, {Gustafsson}  \& {Yong}}{{Mel{\'e}ndez} et~al.}{2009}]{melendez2009}
{Mel{\'e}ndez} J.,  {Asplund} M.,  {Gustafsson} B.,   {Yong} D.,  2009, \mn@doi [\apjl] {10.1088/0004-637X/704/1/L66}, \href {https://ui.adsabs.harvard.edu/abs/2009ApJ...704L..66M} {704, L66}

\bibitem[\protect\citeauthoryear{Meyer, Langer, Mackey, Vel{\'a}zquez  \& Gusdorf}{Meyer et~al.}{2015}]{meyer2015asymmetric}
Meyer D.-A.,  Langer N.,  Mackey J.,  Vel{\'a}zquez P.,   Gusdorf A.,  2015, Monthly Notices of the Royal Astronomical Society, 450, 3080

\bibitem[\protect\citeauthoryear{M{\"u}ller}{M{\"u}ller}{2020}]{muller2020hydrodynamics}
M{\"u}ller B.,  2020, Living Reviews in Computational Astrophysics, 6, 3

\bibitem[\protect\citeauthoryear{Nieva \& Przybilla}{Nieva \& Przybilla}{2012}]{nieva2012present}
Nieva M.-F.,  Przybilla N.,  2012, Astronomy \& Astrophysics, 539, A143

\bibitem[\protect\citeauthoryear{{Nissen, P. E.}}{{Nissen, P. E.}}{2015}]{Nissen1}
{Nissen, P. E.} 2015, \mn@doi [A&A] {10.1051/0004-6361/201526269}, 579, A52

\bibitem[\protect\citeauthoryear{{Nissen, P. E.}}{{Nissen, P. E.}}{2016}]{Nissen2}
{Nissen, P. E.} 2016, \mn@doi [A&A] {10.1051/0004-6361/201628888}, 593, A65

\bibitem[\protect\citeauthoryear{{Nissen, P. E.}, {Christensen-Dalsgaard, J.}, {Mosumgaard, J. R.}, {Silva Aguirre, V.}, {Spitoni, E.}  \& {Verma, K.}}{{Nissen, P. E.} et~al.}{2020}]{Nissen3}
{Nissen, P. E.} {Christensen-Dalsgaard, J.} {Mosumgaard, J. R.} {Silva Aguirre, V.} {Spitoni, E.}  {Verma, K.} 2020, \mn@doi [A&A] {10.1051/0004-6361/202038300}, 640, A81

\bibitem[\protect\citeauthoryear{Ohkubo, Nomoto, Umeda, Yoshida  \& Tsuruta}{Ohkubo et~al.}{2009}]{Ohkubo_2009}
Ohkubo T.,  Nomoto K.,  Umeda H.,  Yoshida N.,   Tsuruta S.,  2009, The Astrophysical Journal, 706, 1184

\bibitem[\protect\citeauthoryear{Orlando et~al.,}{Orlando et~al.}{2025}]{orlando2025tracing}
Orlando S.,  et~al., 2025, Astronomy \& Astrophysics, 699, A305

\bibitem[\protect\citeauthoryear{Proffitt \& Michaud}{Proffitt \& Michaud}{1991}]{proffitt1991gravitational}
Proffitt C.~R.,  Michaud G.,  1991, Astrophysical Journal, Part 1 (ISSN 0004-637X), vol. 380, Oct. 10, 1991, p. 238-250. Research supported by CRSNG and Ministere de l'Education du Quebec., 380, 238

\bibitem[\protect\citeauthoryear{{Roederer} \& {Kirby}}{{Roederer} \& {Kirby}}{2014}]{roederer2014_light}
{Roederer} I.~U.,  {Kirby} E.~N.,  2014, \mn@doi [\mnras] {10.1093/mnras/stu491}, \href {https://ui.adsabs.harvard.edu/abs/2014MNRAS.440.2665R} {440, 2665}

\bibitem[\protect\citeauthoryear{{Roederer} et~al.,}{{Roederer} et~al.}{2014}]{roederer2014_heavy}
{Roederer} I.~U.,  et~al., 2014, \mn@doi [\apj] {10.1088/0004-637X/791/1/32}, \href {https://ui.adsabs.harvard.edu/abs/2014ApJ...791...32R} {791, 32}

\bibitem[\protect\citeauthoryear{Salvadori, Bonifacio, Caffau, Korotin, Andreevsky, Spite  \& Sk{\'u}lad{\'o}ttir}{Salvadori et~al.}{2019}]{10.1093/mnras/stz1464}
Salvadori S.,  Bonifacio P.,  Caffau E.,  Korotin S.,  Andreevsky S.,  Spite M.,   Sk{\'u}lad{\'o}ttir {\'A}.,  2019, \mn@doi [Monthly Notices of the Royal Astronomical Society] {10.1093/mnras/stz1464}, 487, 4261

\bibitem[\protect\citeauthoryear{{Shapley}}{{Shapley}}{1938}]{1938BHarO.908....1S}
{Shapley} H.,  1938, Harvard College Observatory Bulletin, \href {https://ui.adsabs.harvard.edu/abs/1938BHarO.908....1S} {908, 1}

\bibitem[\protect\citeauthoryear{{Sk{\'u}lad{\'o}ttir} et~al.,}{{Sk{\'u}lad{\'o}ttir} et~al.}{2021}]{Skúladóttir_2021}
{Sk{\'u}lad{\'o}ttir} {\'A}.,  et~al., 2021, \mn@doi [\apjl] {10.3847/2041-8213/ac0dc2}, \href {https://ui.adsabs.harvard.edu/abs/2021ApJ...915L..30S} {915, L30}

\bibitem[\protect\citeauthoryear{{Sk{\'u}lad{\'o}ttir}, {Vanni}, {Salvadori}  \& {Lucchesi}}{{Sk{\'u}lad{\'o}ttir} et~al.}{2024a}]{Skúladóttir_2024}
{Sk{\'u}lad{\'o}ttir} {\'A}.,  {Vanni} I.,  {Salvadori} S.,   {Lucchesi} R.,  2024a, \mn@doi [\aap] {10.1051/0004-6361/202346231}, \href {https://ui.adsabs.harvard.edu/abs/2024A&A...681A..44S} {681, A44}

\bibitem[\protect\citeauthoryear{{Sk{\'u}lad{\'o}ttir}, {Koutsouridou}, {Vanni}, {Amarsi}, {Lucchesi}, {Salvadori}  \& {Aguado}}{{Sk{\'u}lad{\'o}ttir} et~al.}{2024b}]{skuladottir2024pair}
{Sk{\'u}lad{\'o}ttir} {\'A}.,  {Koutsouridou} I.,  {Vanni} I.,  {Amarsi} A.~M.,  {Lucchesi} R.,  {Salvadori} S.,   {Aguado} D.~S.,  2024b, \mn@doi [\apjl] {10.3847/2041-8213/ad4b1a}, \href {https://ui.adsabs.harvard.edu/abs/2024ApJ...968L..23S} {968, L23}

\bibitem[\protect\citeauthoryear{Stancliffe \& Glebbeek}{Stancliffe \& Glebbeek}{2008}]{stancliffe2008thermohaline}
Stancliffe R.~J.,  Glebbeek E.,  2008, Monthly Notices of the Royal Astronomical Society, 389, 1828

\bibitem[\protect\citeauthoryear{{Sun} et~al.,}{{Sun} et~al.}{2025}]{Sun2025}
{Sun} Q.,  et~al., 2025, \mn@doi [\apj] {10.3847/1538-4357/ad9924}, \href {https://ui.adsabs.harvard.edu/abs/2025ApJ...980..179S} {980, 179}

\bibitem[\protect\citeauthoryear{Tamborra, Hanke, Janka, Müller, Raffelt  \& Marek}{Tamborra et~al.}{2014}]{Tamborra_2014}
Tamborra I.,  Hanke F.,  Janka H.-T.,  Müller B.,  Raffelt G.~G.,   Marek A.,  2014, \mn@doi [The Astrophysical Journal] {10.1088/0004-637X/792/2/96}, 792, 96

\bibitem[\protect\citeauthoryear{Thibodeaux, Ji, Cerny, Kirby  \& Simon}{Thibodeaux et~al.}{2024}]{Thibodeaux2024LAMOST}
Thibodeaux P.,  Ji A.~P.,  Cerny W.,  Kirby E.~N.,   Simon J.~D.,  2024, \mn@doi [The Open Journal of Astrophysics] {10.33232/001c.122335}, 7

\bibitem[\protect\citeauthoryear{Tinsley}{Tinsley}{1979}]{tinsley1979stellar}
Tinsley B.~M.,  1979, Astrophysical Journal, Part 1, vol. 229, May 1, 1979, p. 1046-1056. Research supported by the Alfred P. Sloan Foundation, 229, 1046

\bibitem[\protect\citeauthoryear{Tominaga, Umeda  \& Nomoto}{Tominaga et~al.}{2007}]{Tominaga_2007}
Tominaga N.,  Umeda H.,   Nomoto K.,  2007, \mn@doi [The Astrophysical Journal] {10.1086/513063}, 660, 516

\bibitem[\protect\citeauthoryear{Vartanyan, Burrows, Radice, Skinner  \& Dolence}{Vartanyan et~al.}{2019}]{vartanyan2019successful}
Vartanyan D.,  Burrows A.,  Radice D.,  Skinner M.~A.,   Dolence J.,  2019, Monthly Notices of the Royal Astronomical Society, 482, 351

\bibitem[\protect\citeauthoryear{Vartanyan, Tsang, Kasen, Burrows, Wang  \& Teryoshin}{Vartanyan et~al.}{2025}]{vartanyan20253d}
Vartanyan D.,  Tsang B. T.-H.,  Kasen D.,  Burrows A.,  Wang T.,   Teryoshin L.,  2025, The Astrophysical Journal, 982, 9

\bibitem[\protect\citeauthoryear{{Venn} et~al.,}{{Venn} et~al.}{2012}]{Venn}
{Venn} K.~A.,  et~al., 2012, \mn@doi [\apj] {10.1088/0004-637X/751/2/102}, \href {https://ui.adsabs.harvard.edu/abs/2012ApJ...751..102V} {751, 102}

\bibitem[\protect\citeauthoryear{{Wang} \& {Burrows}}{{Wang} \& {Burrows}}{2024}]{Wang_and_Burrows_2024}
{Wang} T.,  {Burrows} A.,  2024, \mn@doi [\apj] {10.3847/1538-4357/ad12b8}, \href {https://ui.adsabs.harvard.edu/abs/2024ApJ...962...71W} {962, 71}

\bibitem[\protect\citeauthoryear{Wasserburg, Busso, Gallino  \& Nollett}{Wasserburg et~al.}{2006}]{wasserburg2006short}
Wasserburg G.,  Busso M.,  Gallino R.,   Nollett K.~M.,  2006, Nuclear Physics A, 777, 5

\bibitem[\protect\citeauthoryear{White \& Springel}{White \& Springel}{2000}]{white1999}
White S. D.~M.,  Springel V.,  2000, in Weiss A.,  Abel T.~G.,   Hill V.,  eds, The First Stars. Springer Berlin Heidelberg, Berlin, Heidelberg, pp 327--335

\bibitem[\protect\citeauthoryear{Xing et~al.,}{Xing et~al.}{2023}]{xing2023metal}
Xing Q.-F.,  et~al., 2023, Nature, 618, 712

\bibitem[\protect\citeauthoryear{Yong et~al.,}{Yong et~al.}{2021}]{yong2021r}
Yong D.,  et~al., 2021, Nature, 595, 223

\bibitem[\protect\citeauthoryear{van Baal, Jerkstrand, Wongwathanarat  \& Janka}{van Baal et~al.}{2023}]{van2023modelling}
van Baal B.~F.,  Jerkstrand A.,  Wongwathanarat A.,   Janka H.-T.,  2023, Monthly Notices of the Royal Astronomical Society, 523, 954

\makeatother
\end{thebibliography}



\appendix

\section{Additional fits and information}

\twocolumn
\begin{figure}
    \centering
    \includegraphics[width=1\linewidth]{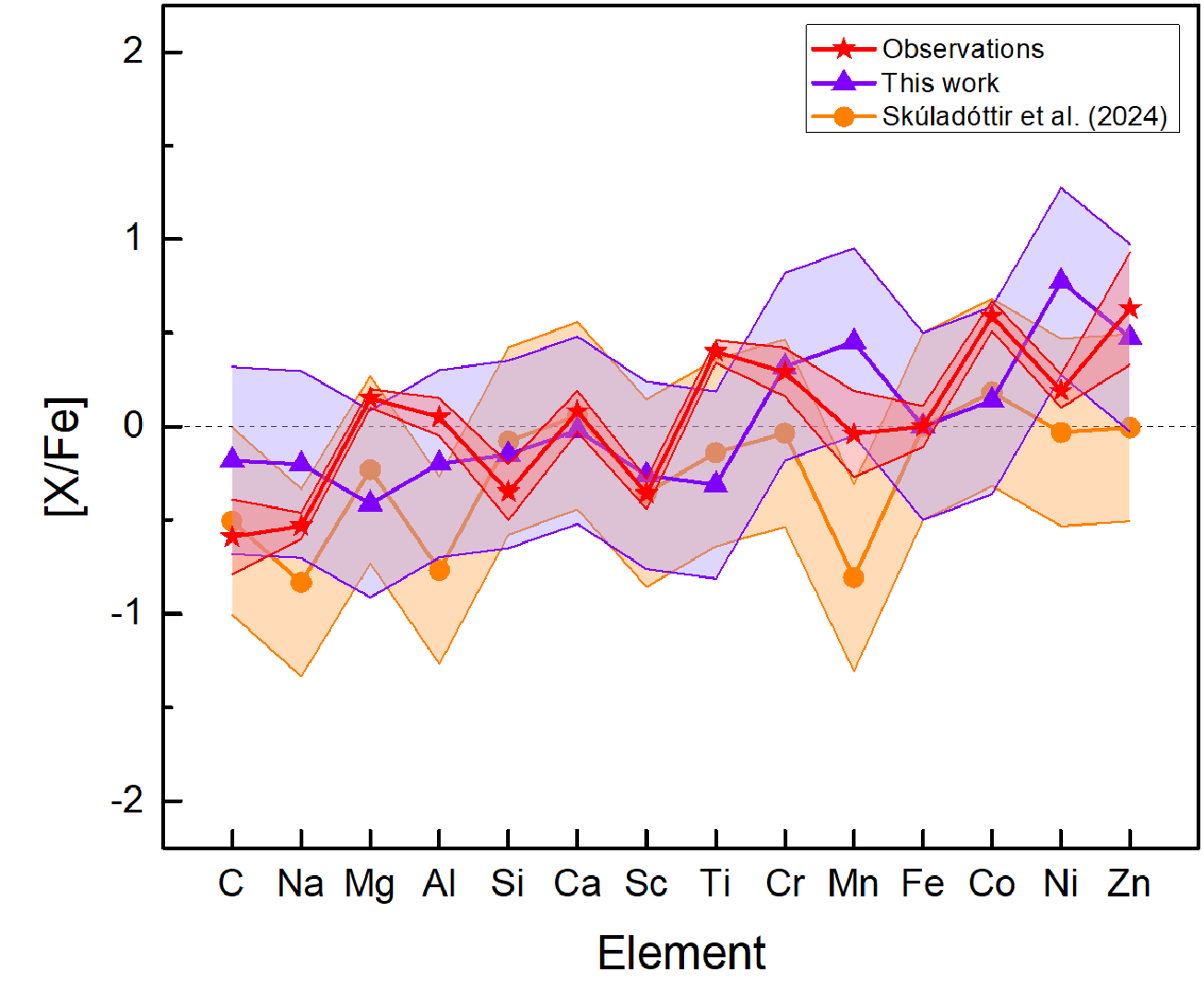}
    \caption{Same as Fig. \ref{fig:main_fit} but for a 15 \(\textup{M}_\odot\) model.}
    \label{fig:15_SM}
\end{figure}

\begin{figure}
    \centering
    \includegraphics[width=1\linewidth]{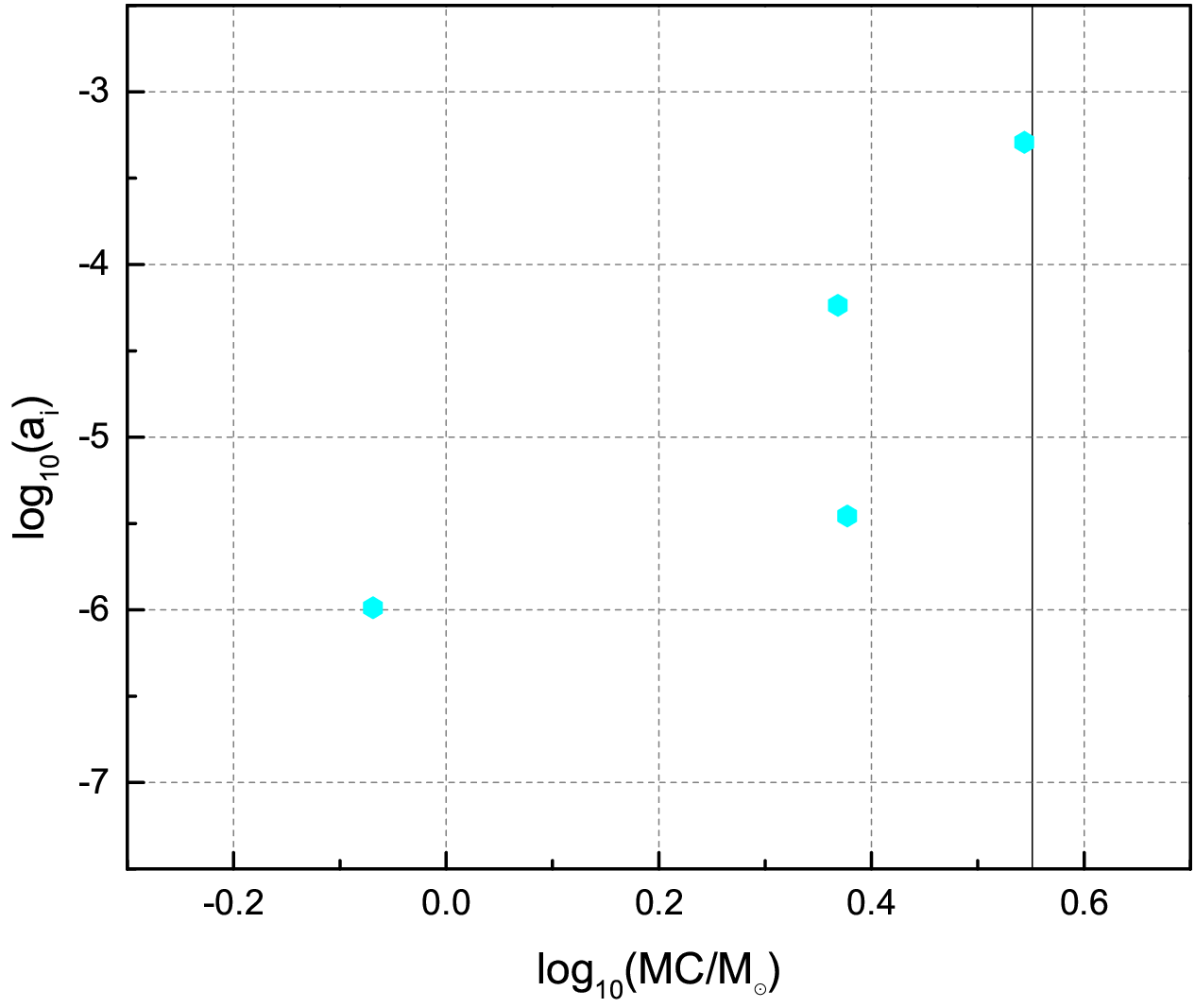}
    \caption{Same as Fig. \ref{fig:coeff} but for a 15 \(\textup{M}_\odot\) model.}
    \label{fig:15_SM_coeff}
\end{figure}

\begin{figure}
    \centering
    \includegraphics[width=1\linewidth]{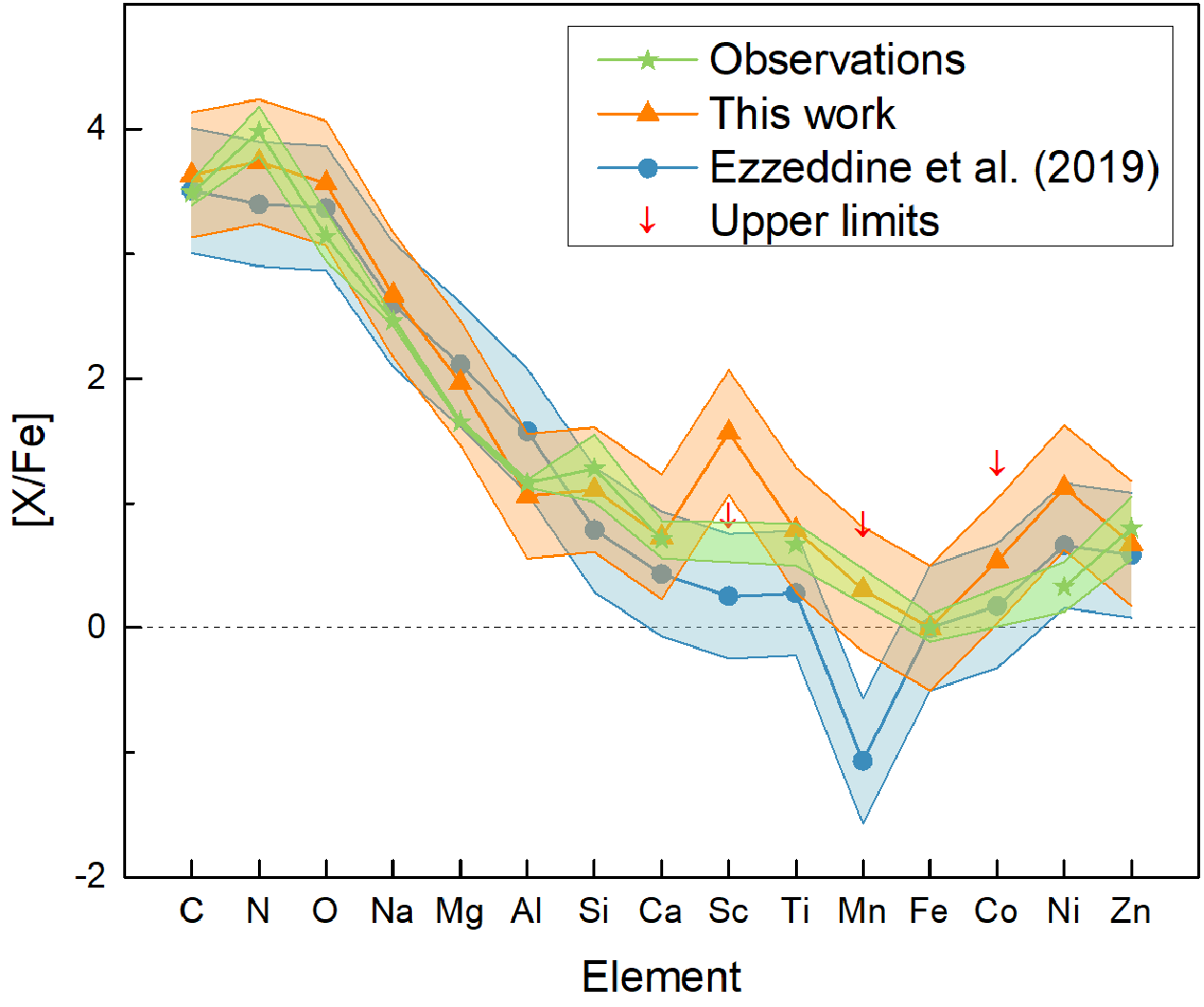}
    \caption{Same as Fig. \ref{fig:HE_main_fit} but with another set of coefficients for the same model and shells and with upper limits. Coefficients presented in Fig. \ref{fig:HE_ul_coeff}. ($\chi^2 \sim 4.05$)}
    \label{fig:HE_ul_fit}
\end{figure}

\begin{figure}
    \centering
    \includegraphics[width=1\linewidth]{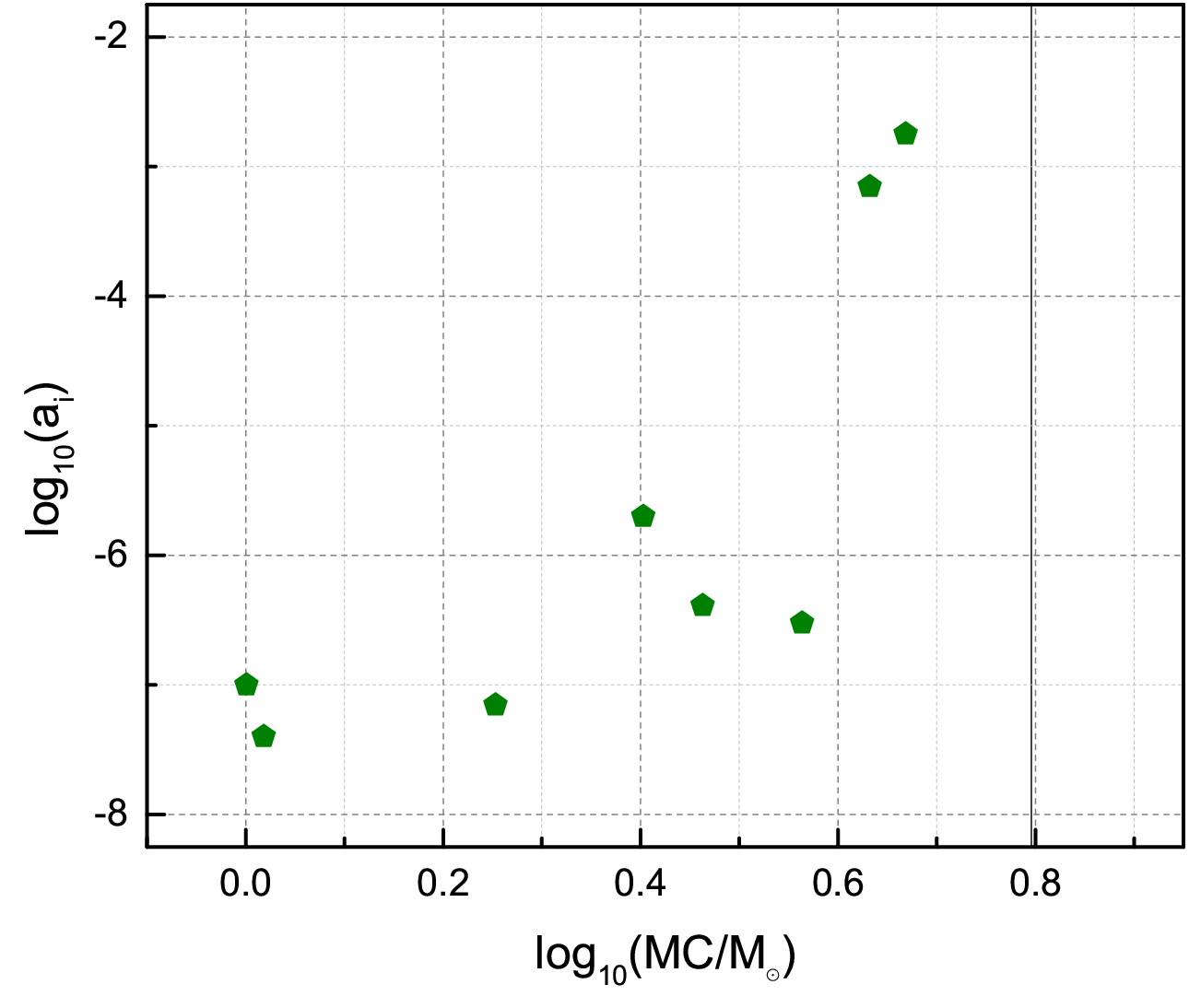}
    \caption{Same as Fig. \ref{fig:coeff} but for HE 1327-2326.}
    \label{fig:HE_ul_coeff}
\end{figure}

\begin{figure}
    \centering
    \includegraphics[width=1\linewidth]{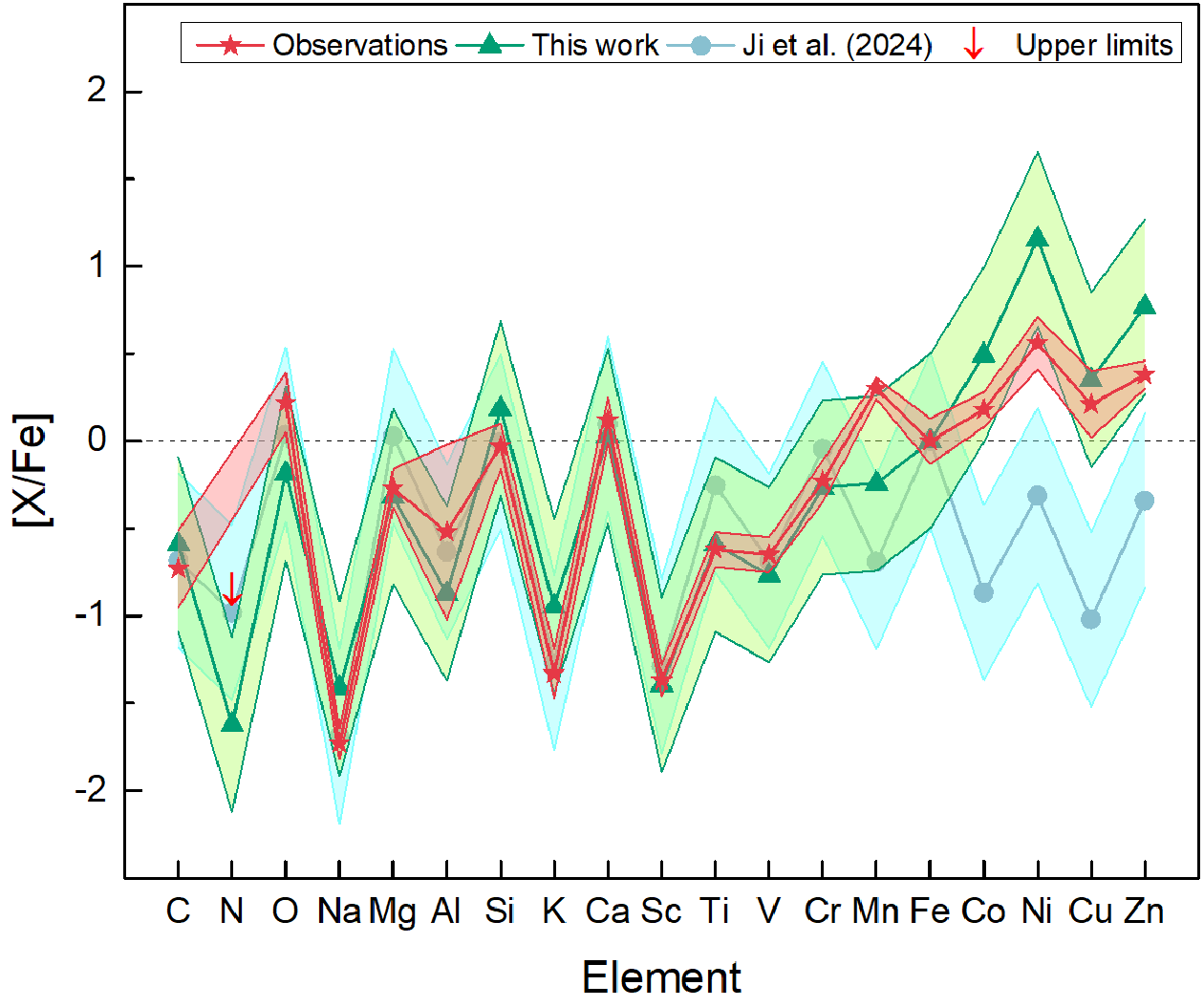}
    \caption{Same as Fig. \ref{fig:80_SM_main_fit} but using another set of coefficients and shells for the same model and with upper limits. Coefficients presented in Fig. \ref{fig:80_SM_ul_coeff}. ($\chi^2 \sim 5.61$)}
    \label{fig:80_SM_ul_fit}
\end{figure}

\begin{figure}
    \centering
    \includegraphics[width=1\linewidth]{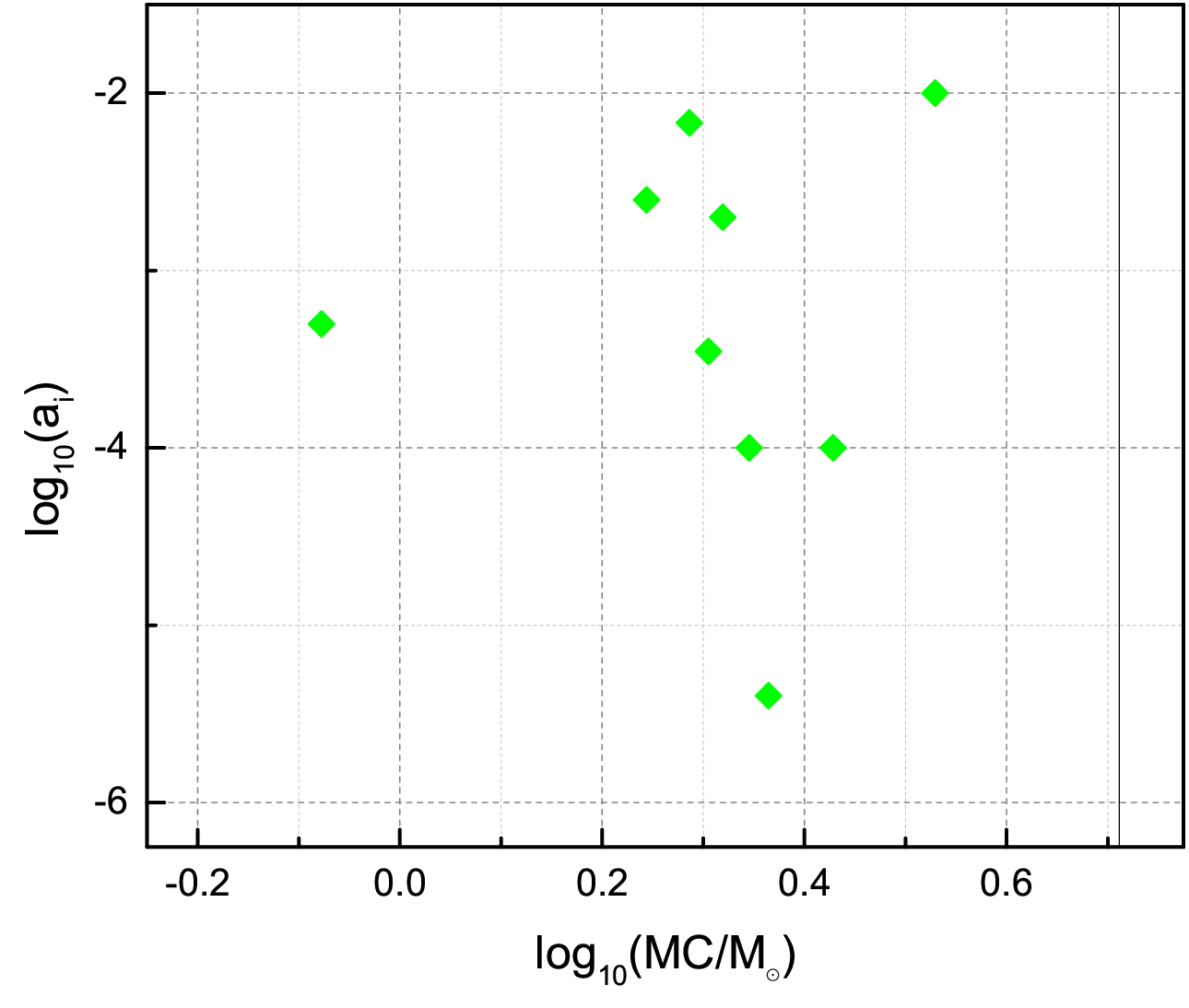}
    \caption{Same as Fig. \ref{fig:coeff} but for J0931+0038.}
    \label{fig:80_SM_ul_coeff}
\end{figure}

\begin{figure}
    \centering
    \includegraphics[width=1\linewidth]{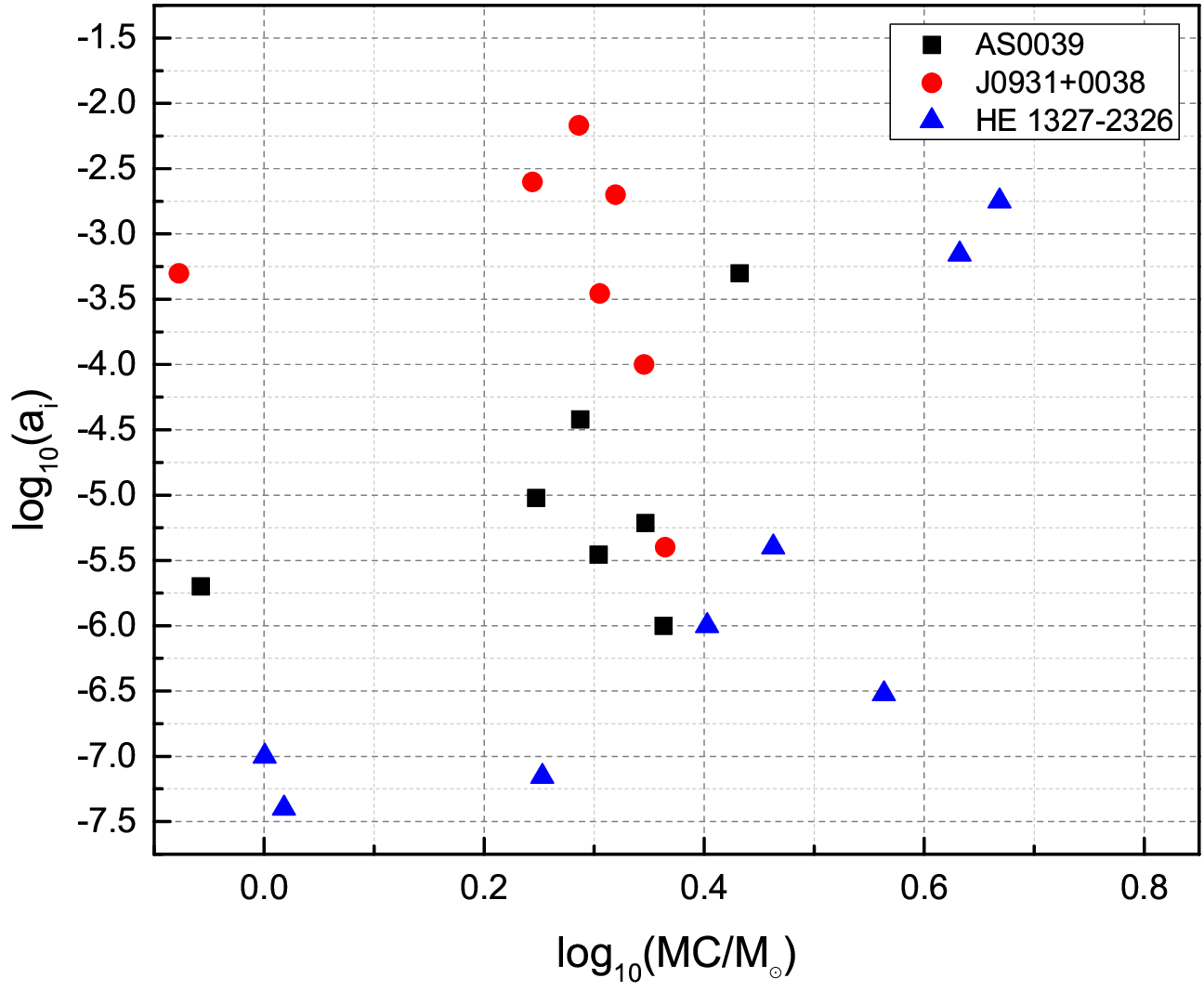}
    \caption{All the coefficients used for the main fits. The Durbin Watson (DW) test statistic shows a first order positive autocorrelation between the coefficients to a high significance level. The value we obtain for the DW test is 1.14, which is below 1.24 (the significance threshold for 95\% confidence in our case of n = 22 and k = 1).}
    \label{fig:all_coeff}
\end{figure}

\twocolumn
\begin{table}
    \centering
    \caption{Fitting and residual values for the 13 \(\textup{M}_\odot\) model for AS0039.}
    \label{tab:residuals}
    \begin{tabular}{|c|c|c|c|c|c|c|}
        \hline
        Element & \text{[X/Fe]\textsubscript{NLTE (obs)}} & $\delta\textsubscript{[X/Fe] (obs)}$ & \text{[X/Fe]\textsubscript{\cite{Skúladóttir_2024}}} & \text{Residuals\textsubscript{\cite{Skúladóttir_2024}}} & \text{[X/Fe]\textsubscript{(This work)}} & \text{Residuals\textsubscript{(This work)}} \\
        \hline
        C  & -0.59 & 0.20 & -0.51 & -0.08 & -0.60 & 0.01 \\
        Na & -0.53 & 0.07 & -0.83 & 0.30 & -0.35 & -0.18 \\
        Mg & 0.15 & 0.05 & -0.23 & 0.38 & -0.03 & 0.18 \\
        Al & 0.05 & 0.10 & -0.77 & 0.82 & 0.10 & -0.05 \\
        Si & -0.35 & 0.15 & -0.08 & -0.27 & 0.05 & -0.40 \\
        Ca & 0.08 & 0.11 & 0.06 & 0.02 & 0.16 & -0.08 \\
        Sc & -0.36 & 0.08 & -0.36 & -0.00 & -0.43 & 0.07 \\
        Ti & 0.40 & 0.06 & -0.14 & 0.54 & -0.31 & 0.71 \\
        Cr & 0.29 & 0.13 & -0.04 & 0.33 & 0.13 & 0.16 \\
        Mn & -0.04 & 0.23 & -0.81 & 0.77 & -0.01 & -0.03 \\
        Fe & 0.00 & 0.11 & 0.00 & 0.00 & 0.00 & 0.00 \\
        Co & 0.59 & 0.08 & 0.18 & 0.41 & 0.08 & 0.51 \\
        Ni & 0.19 & 0.09 & -0.03 & 0.22 & 0.56 & -0.37 \\
        Zn & 0.63 & 0.30 & -0.01 & 0.64 & -0.47 & 1.10 \\
        \hline
    \end{tabular}
\end{table}

\begin{table}
    \centering
    \caption{Percentage contribution of each shell to each element in the 13 \(\textup{M}_\odot\) model fit for AS0039.}
    \label{tab:element_composition}
    \begin{tabular}{|c|c|c|c|c|c|c|c|c|}
    \hline
    Element & Shell 1 & Shell 2 & Shell 3 & Shell 4 & Shell 5 & Shell 6 & Shell 7 & Total \\ \hline
        C  & 2.8101E-03 & 3.1758E-03 & 9.5215E-06 & 8.9085E-04 & 1.5543E-03 & 1.3837E-01 & 9.9853E+01 & 100 \\
        Na & 1.6166E-09 & 1.7627E-04 & 1.0804E-09 & 6.1662E-05 & 1.2073E-04 & 1.2893E-01 & 9.9871E+01 & 100 \\
        Mg & 2.6692E-05 & 3.4821E-04 & 6.6480E-04 & 1.6905E-03 & 6.5472E-03 & 4.0402E-01 & 9.9587E+01 & 100 \\
        Al & 1.0841E-05 & 1.0205E-03 & 1.5227E-06 & 2.7130E-04 & 1.6920E-03 & 5.0425E-01 & 9.9493E+01 & 100 \\
        Si & 2.7288E-04 & 2.7783E-01 & 3.9506E+01 & 1.4921E+01 & 8.3202E+00 & 2.2392E+00 & 3.4735E+01 & 100 \\
        Ca & 1.3127E-01 & 2.7690E+00 & 6.3964E+01 & 1.6593E+01 & 1.5038E+01 & 1.5038E+00 & 5.1267E-04 & 100 \\
        Sc & 8.2718E-01 & 8.8164E+01 & 8.5127E-01 & 2.1392E+00 & 6.3775E+00 & 6.9105E-01 & 9.5012E-01 & 100 \\
        Ti & 6.9920E+00 & 1.8887E+01 & 6.0451E+01 & 5.0033E+00 & 8.0066E+00 & 5.6921E-01 & 9.1096E-02 & 100 \\
        Cr & 1.5474E+00 & 1.2237E+01 & 8.0654E+01 & 4.7638E+00 & 7.6537E-01 & 2.6756E-02 & 5.6316E-03 & 100 \\
        Mn & 1.0760E+01 & 8.6933E+00 & 7.3073E+01 & 7.2004E+00 & 2.6891E-01 & 2.7374E-03 & 1.1667E-03 & 100 \\
        Fe & 3.0803E+01 & 3.9944E+01 & 2.7404E+01 & 1.7877E+00 & 6.0448E-02 & 4.3450E-04 & 5.6233E-05 & 100 \\
        Co & 8.2263E+01 & 1.7662E+01 & 9.2291E-03 & 4.4512E-02 & 1.7327E-02 & 2.0396E-04 & 3.6401E-03 & 100 \\
        Ni & 8.7693E+01 & 1.0420E+01 & 1.5382E+00 & 3.3031E-01 & 1.7761E-02 & 9.8574E-05 & 3.9144E-04 & 100 \\
        Zn & 6.5677E+01 & 3.4161E+01 & 1.1356E-08 & 3.6988E-06 & 1.5012E-04 & 1.1031E-03 & 1.6034E-01 & 100 \\
    \hline
    \end{tabular}
\end{table}


\bsp	
\label{lastpage}
\end{document}